\documentclass[%
reprint,%
aps,%
prl,%
groupedaddress,%
superscriptaddress,%
assume,%
amsmath,%
floatfix%,
]{revtex4-2}

\usepackage{graphicx}
\usepackage[
  colorlinks=true,
  urlcolor=blue,
  linkcolor=blue,
  citecolor=blue
]{hyperref}
\usepackage{siunitx}
\usepackage{braket}
\usepackage{gensymb}
\usepackage{amssymb}

\begin{document}
\title{Space-Time Hopfion Crystals}
\author{Wenbo Lin}\email{lin.w.ab@m.titech.ac.jp}
\affiliation{Institute of Innovative Research, Tokyo Institute of Technology, 2-12-1 Ookayama, Merugo, Tokyo 152-8550, Japan}
%\affiliation{Research Center for Advanced Science and Technology, The University of Tokyo, 4-6-1 Komaba, Meguro, Tokyo 153-8904, Japan}
\author{Nilo Mata-Cervera}
\affiliation{Centre for Disruptive Photonic Technologies, School of Physical and Mathematical Sciences \& The Photonics Institute, Nanyang Technological University, Singapore 637371, Singapore}
\author{Yasutomo Ota}
\affiliation{Department of Applied Physics and Physico-Informatics, Faculty of Science and Technology, Keio University, Yokohama, Kanagawa 223-8522, Japan}
\author{Yijie Shen}\email{yijie.shen@ntu.edu.sg}
\affiliation{Centre for Disruptive Photonic Technologies, School of Physical and Mathematical Sciences \& The Photonics Institute, Nanyang Technological University, Singapore 637371, Singapore}
\affiliation{School of Electrical and Electronic Engineering, Nanyang Technological University, Singapore 639798, Singapore}
\author{Satoshi Iwamoto}\email{iwamoto@iis.u-tokyo.ac.jp}
\affiliation{Research Center for Advanced Science and Technology, The University of Tokyo, 4-6-1 Komaba, Meguro, Tokyo 153-8904, Japan}
\affiliation{Institute of Industrial Science, The University of Tokyo, 4-6-1 Komaba, Meguro, Tokyo 153-8505, Japan}

\date{\today}

\begin{abstract}
Hopfions, higher-dimensional topological quasiparticles with sophisticated 3D knotted spin textures discovered in condensed matter and photonic systems, show promise in high-density data storage and transfer. Here we present crystalline structures of hopfions lying in space-time constructed by spatiotemporally structured light. A practical methodology using bichromatic structured light beams or dipole arrays to assemble 1D and higher dimensional hopfion lattices is proposed and a technique for tailoring topological orders is elucidated. The birth of photonic hopfion crystals heralds a new era in high-dimensional, condensed, and robust topological information processing.
\end{abstract}

\maketitle

Particle-like localized spin textures with complex topological structures have recently emerged as potential information carriers in solid-state magnetic materials~\cite{bogdanov2020physical,gobel2021beyond,bernevig2022progress,han2022high,chen2024all} and structured optical fields~\cite{shen2024optical,he2022towards,wan2023ultra,lin2021microcavity,kerridge2024optical}. As the family of 2D spin textures topologically classified by the homotopy group $\pi_2(S^2)=\mathbb{Z}$, skyrmions and their complex crystalline structures, i.e., skyrmion crystals or lattices, have already revolutionized high-density 2D data storage and transfer technologies~\cite{fert2017magnetic,yu2018transformation,foster2019two,tang2021magnetic,hayami2021phase,tsesses2018optical,du2019deep,lei2021photonic,shen2022generation,ornelas2024non}.

In contrast to these 2D textures, hopfions, holding sophisticated 3D toroidal knotted topology fulfilling $\pi_3(S^2)=\mathbb{Z}$, have recently been discovered in magnetic materials~\cite{zheng2023hopfion,liu2020three,liu2022emergent,kent2021creation,rybakov2022magnetic} and photonic fields~\cite{sugic2021particle,shen2023topological,wang2023photonic}. Hopfions can be considered as closed twisted skyrmion strings and promise advancements in processing higher-dimensional information. However, existing methods can only generate isolated hopfions, and no methodology for incarnating crystalline structures of hopfions is known so far.

In this Letter, we demonstrate that spatiotemporal light fields, which have recently been utilized to construct topological textures such as skyrmions and scalar vortex rings~\cite{shen2023roadmap,shen2021supertoroidal,zdagkas2022observation,wan2022toroidal,wan2022scalar,zhan2024cylindrical,shen2024nondiffracting}, can also be used to embody 3D topologies of hopfion crystals, filling the gap in the topological family. The hopfion textures are constructed by the polarization pseudospin vector defined in a bichromatic light field. We first show how to shape the polychromatic light field to form 1D hopfion lattices periodically repeated in time. Then, we extend our model into 3D hopfion crystals with simultaneously spatial and temporal periodicities.

\begin{figure}[b]
	\centering
	\includegraphics{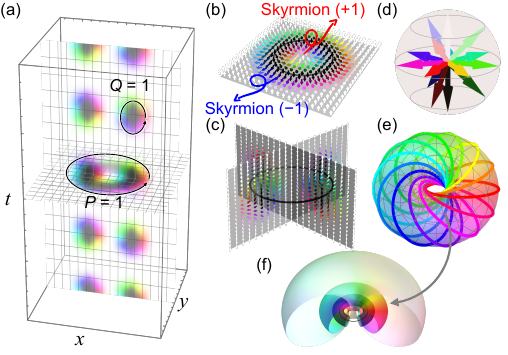}
	\caption{%
        (a)~Concept of a 1D space-time hopfion crystal.
        (b)~Spin textures on $x$-$y$ transverse cross-section and (c)~that on selective longitudinal cross-sections through the center of a unit hopfion lattice.
        (d)~The arrow indicator of all states of spin on a parametric sphere $S^2$.
        (e)~The isospin lines of selected spins on the equator of the parametric sphere form a torus-knot configuration.
        (f)~All isospin lines of a hopfion lattice show a layer-by-layer torus-knot fibration configuration.
    }
	\label{f1}
\end{figure}
Firstly, we discuss one-dimensional hopfion crystals. Figure~\ref{f1}(a) showcases a fundamental hopfion crystal of first-order. In this configuration, the time-variant 2D spin texture at $(x,y)$ plane periodically repeats along time $t$, forming a chain of 3D hopfion texture. Each unit cell possesses a localized 3D hopfion spin texture exactly fulfilling the Hopf map $\pi_3(S^2)=1$ with a stereographic projection from $S^3$ to $\mathbb{R}^3$ ($S^3\backslash\left\{ 0 \right\}\cong\mathbb{R}^3$). The Hopf map encompasses skyrmion maps ($S^2\to\mathbb{R}^2$) within its subspaces. Figure~\ref{f1}(b) depicts the transverse cross-section through the center of a hopfion unit cell, showcasing a skyrmionium texture: two radially nested skyrmions with opposite polarity (marked by ``$\pm1$''). Another cross-section along the $t$-axis includes a pair of skyrmions with opposite polarity, as depicted in Fig.~\ref{f1}(c). This arrangement can be interpreted by bending and twisting a skyrmion tube along the boundary of the radial nesting, indicated by black spin-down circles in Figs.~\ref{f1}(b--c).

For higher order cases, the Hopf number is given by the product of two skyrmion numbers $Q_H=P\times Q$, where $P$ and $Q$ respectively characterize $(x,y)$ and $(x,t)$ skyrmions. A unit skyrmion can be rendered as a full mapping of the spin vectors from a 2-sphere to a 2D plane ($S^2\to\mathbb{R}^2$), as depicted in Fig.~\ref{f1}(d) with a colormap representing a unit spin vector $\boldsymbol{s} = (s_x,s_y,s_z)$ used in this paper. Here, lightness (brightness and darkness) corresponds to the value of the $s_z$ component (up and down), while hue represents the azimuth of the transverse $(s_x,s_y)$ component. In a skyrmion region, all spin states (corresponding to points on the parametric 2-sphere) are involved and the full lightness-hue colors are observed. In contrast, the 3D spin texture of a hopfion cell fulfills a higher-dimensional map $S^3 \to \mathbb{R}^3$, a map from a 4D hypersphere to a 3D volume. This implies that a spin state is mapped to no longer a point but a closed loop in real space to introduce an additional dimension. A set of loops as isospin lines mapped from selective spins on the equator of the parametric sphere forms a torus-knot configuration, where the knotting number of each loop also relates to hopfion topology: $P$ counts how many times a loop goes through the hole, and $Q$ how many time it goes round the torus, as illustrated in Fig.\ref{f1}(e) for the case of a fundamental hopfion with $Q_H=P=Q=1$. The full mapping of spins through all the latitudes from south to north poles (spin down to up) forms a configuration of layer-by-layer torus-knots from the central spin-down ring to the spin-up unit cell boundary, known as Hopf fibration, as depicted in Fig.~\ref{f1}(f).

\begin{figure}
	\centering
	\includegraphics{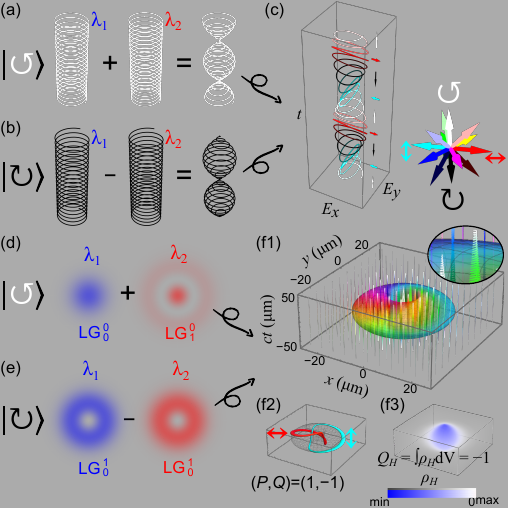}
	\caption{%
        (a)~Spatiotemporal trajectory of bichromatic $\ket{\circlearrowleft}$-polarized electric field and (b)~$\ket{\circlearrowright}$-polarized variant with a $\pi$-phase shift for one color, causing differing beat timings.
        (c)~Their superposition exhibits a time-varying pseudospin profile.
        Spatial modes of (d)~$\ket{\circlearrowleft}$- and (e)~$\ket{\circlearrowright}$-polarized components for hopfionic light field synthesis. Due to diverse spatial modes, (f1)~polarization state from their superposition varies spatially (displayed by colored lines like (c) as partially zoomed in the inset, high-resolution images can be found in SI). If one pseudospin component remains fixed, its distribution forms a torus (shown for $s_z=0$), while (f2)~spatial distribution of any pseudospin forms a loop wrapping around a torus, with loops of distinct pseudospins possessing a constant linking number $Q_H=P \times Q$. This number can be further verified by evaluating (f3) Hopf density and integrating it over volume.
    } 
	\label{f2}
\end{figure}
To embody such a hopfion crystal in an optical field, a time-dependent pseudospin vector must be defined. Here we exploit the time variation of the electric field vector created by a coherent light wave with two colors. When the ratio of two wavelengths $\lambda_1$ and $\lambda_2$ is a rational number, the trajectory of the electric field vector of such a two-color light field draws a Lissajous curve, known as a Lissajous beam~\cite{zhu2024directional}. Particularly, we assume a paraxial bichromatic field and the electric field parallel to the wave vector is negligible. The two distinct wavelengths let the total electric field $\boldsymbol{E}(t) = \boldsymbol{E}_1 e^{-i\frac{2 \pi}{\lambda_1}ct} + \boldsymbol{E}_2 e^{-i\frac{2 \pi}{\lambda_2}ct}$ varies on time, where $c$ is the speed of light. When $\lambda_1 / \lambda_2$ is a rational number, the total electric field beats with a certain temporal period, as depicted for $\ket{\circlearrowleft}$ polarization in Fig.~\ref{f2}(a) ($\ket{\circlearrowleft}$ and $\ket{\circlearrowright}$ refer to the left- and right-handed circular polarizations, respectively). By superimposing another bichromatic field with different polarization and phase, such as shown in Fig.~\ref{f2}(b), we can obtain a field with a continuously varying polarization ellipse, as rendered in Fig.~\ref{f2}(c). We define temporal pseudospin vector to characterize such time-varying polarization states as
\begin{equation}
   \boldsymbol{s}(t)= \begin{pmatrix} s_x(t)\\ s_y(t)\\ s_z(t) \end{pmatrix} = \frac{1}{\boldsymbol{E}^\ast(t)\boldsymbol{E}(t)}\begin{pmatrix}
        \boldsymbol{E}^\ast(t)\sigma_3\boldsymbol{E}(t)\\
        \boldsymbol{E}^\ast(t)\sigma_1\boldsymbol{E}(t)\\
        \boldsymbol{E}^\ast(t)\sigma_2\boldsymbol{E}(t)
    \end{pmatrix},
    \label{eq1}
\end{equation}
where $\sigma_1$, $\sigma_2$, and $\sigma_3$ are Pauli matrices. This is an extension of the ``normalized'' Stokes vector for the total electric field vector of a bichromatic field. The pseudospin vector approximates the instantaneous Stokes vector of the polarization ellipse if the two wavelengths are close and the polarization ellipse varies slowly to the wavelengths.

To sculpt the vector structure and construct hopfionic topology spatiotemporally, we further modulate the spatial properties of these temporally periodic light fields by assigning different spatial modes to each circular polarization and wavelength component. Note that although we use circular polarizations, any pair of orthogonal polarization states can serve the same purpose. We start with the superposition of two Laguerre-Gaussian (LG) modes $\mathrm{LG}_0^0$ and $\mathrm{LG}_1^0$ ($\mathrm{LG}_p^\ell$ refers to the mode with azimuthal index $\ell$ and radial index $p$) with two wavelengths $\lambda_1$ and $\lambda_2$ for $\ket{\circlearrowleft}$ polarization. This guarantees that the pseudospin points upwards at the optical axis and infinitely far away from it, as rendered in Fig.~\ref{f2}(d).
To realize the hopfion structure, generating a spin-down circle in spacetime is imperative. Since $\mathrm{LG}_0^0$ maintains phase coherence across the transverse plane while $\mathrm{LG}_1^0$ undergoes a $\pi$-phase shift beyond a certain radius, these two modes construct a ring of zero intensity due to destructive interference, thereby facilitating the formation of a spin-down circle by further superimposing $\mathrm{LG}_0^1$ in the orthogonal polarization $\ket{\circlearrowright}$. If the wavelengths of $\mathrm{LG}_0^0$ and $\mathrm{LG}_1^0$ differ, such destructive interference will occur periodically and instantaneously. Another prerequisite is to induce a moment of uniform spin-up to satisfy the hopfion's boundary condition on the temporal axis. This is achieved by also utilizing $\mathrm{LG}_0^1$ with two wavelengths and superimposing them with a $\pi$-phase shift to counterbeat the $\ket{\circlearrowleft}$ polarization, as illustrated in Fig.~\ref{f2}(e). The resultant spatiotemporal hopfionic field can be expressed as:
\begin{equation}
\begin{split}
    \boldsymbol{E}(\mathbf{r},t)=&
    \left[\mathrm{LG}_0^0(\mathbf{r},t|\lambda_1)+\mathrm{LG}_1^0(\mathbf{r},t|\lambda_2)\right]\ket{\circlearrowleft}\\
    &+ \left[\mathrm{LG}_0^1(\mathbf{r},t|\lambda_1)-\mathrm{LG}_0^1(\mathbf{r},t|\lambda_2)\right]\ket{\circlearrowright}.
\end{split}
\label{eq2}
\end{equation}
where in principle the spatial coordinate $\mathbf{r}=(x,y,z)$ can include transverse plane $(x,y)$ and propagation axis $z$, while, as a proof of concept we only consider $z=0$ (beam waist) and observe the topological textures in 3D space-time $(x,y,t)$, noted as $(\mathbf{r}_{\perp},t)$ in brief. 
Figure~\ref{f2}(f1) shows a single period of the spatiotemporal structure of linear polarization generated at the beam waist (waist radius \SI{10}{\micro\meter}) when $\lambda_1 = 1$\,\si{\micro\meter} and $\lambda_2 = 1/1.01$\,\si{\micro\meter}. The wavelength ratio $100/101$ gives a temporal beat period of $ct=\SI{100}{\micro\meter}$. A collection of isospin loops forms a torus (Fig.~\ref{f2}(f1)), and two distinct isospin loops are linked once, as depicted in Fig.~\ref{f2}(f2).
We characterize the topology of the space-time pseudospin field in Fig.~\ref{f2}(f1) by calculating its topological invariant of Hopf number~\cite{Hopf_invariant_Whitehead,Hopf_calculation_Fourier}. Figure~\ref{f2}(f3) depicts the computed Hopf density $\rho_{H}(\mathbf{r}_{\perp},t)$ in a single temporal period (see Supplementary Information SI for calculation details). The integral value is near-unity, exhibiting a remarkable quality of the first-order hopfion topology.

\begin{figure}
	\centering
	\includegraphics{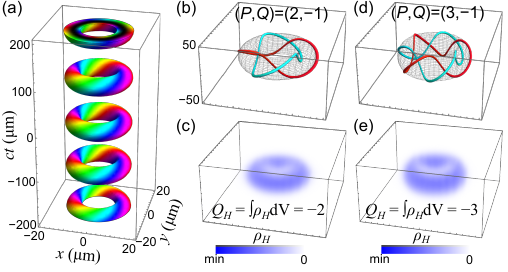}
	\caption{%
        Realization of high order hopfion crystals: (a)~pseudospin textures in space-time for $s_z\leq0$, representing several temporal periods. The lightness represents the value of $s_z$ from $-1$ to $0$ and the color hue represents $h=\arctan{\left(s_y/s_x\right)}$. Isospin curves for $(P,Q)=(2,-1)$ (b) and $(P,Q)=(3,-1)$ (d) in a single temporal period and their corresponding Hopf densities (c) and (e) respectively.
    } 
	\label{f3}
\end{figure}

\begin{figure*}[t]
	\centering
	\includegraphics{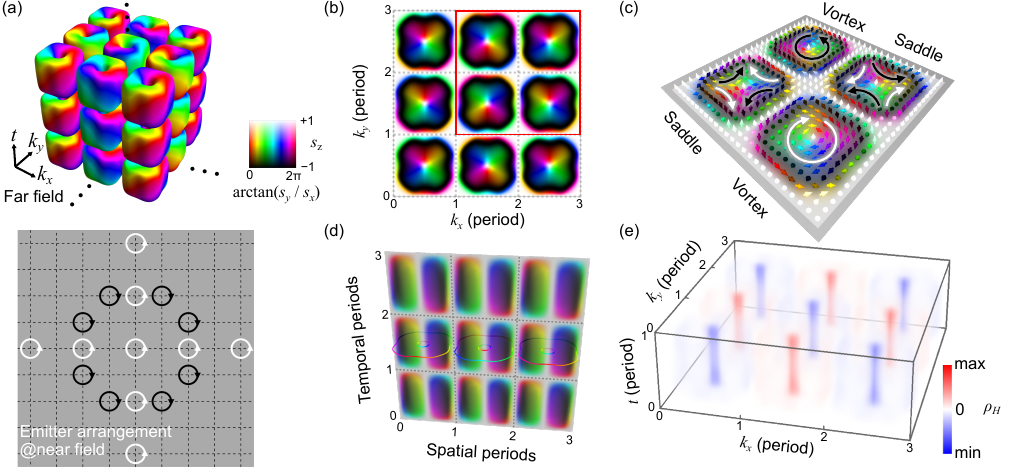}
	\caption{%
        (a)~Concept of 3D space-time hopfion crystal (top) and its schematic generation setup (bottom).
        (b)~A scalar plot of the pseudospin field in the transversal plane. The red box highlights the unit supercell of the crystal.
        (c)~Vector plot of the pseudospin field in the unit supercell, where the white and black arrows show clockwise and counter-clockwise rotation of the vectors that yield vortex and saddle textures. (d)~Far-field pseudospin vector field in the $k_y-t$ plane.
        (e)~Hopf density $\rho_{H}(k_x,k_y,t)$ in a single temporal period. 
    }
	\label{f4}
\end{figure*}

Higher-order hopfion crystals with unit cells featuring arbitrary Hopf numbers can be realized if one of the two skyrmion numbers $(P, Q)$ can be controlled. Here we discuss a scheme for tailoring $P$, the skyrmion number characterizes the skyrmionium in the transverse plane. Similar to the generation of higher-order skyrmions using higher-order OAM modes~\cite{lin2021microcavity,shen2022generation}, a higher-order skyrmionium with an order of $P=\ell$ can be formed by superimposing $\ket{\circlearrowright}$-polarized $\mathrm{LG}_0^\ell$ mode and $\ket{\circlearrowleft}$-polarized $\mathrm{LG}_p^0$ modes. Note that $P=-\ell$ is obtained by swapping $\ket{\circlearrowright}$ and $\ket{\circlearrowleft}$. As in the first-order case, the spin-up must be satisfied at the center of the transverse plane and at infinities/boundaries. Since the $\mathrm{LG}_0^\ell$ mode is characterized by a polynomial factor of $O\left(r^{\left| \ell \right|}\right)$ for the radial coordinate $r$, while the polynomial part of $\mathrm{LG}_p^0$ is $O\left(r^{2p}\right)$, at least two $\mathrm{LG}_p^0$ modes respectively satisfying $p < \left| \ell \right|/2$ and $p > \left| \ell \right|/2$ are required to fulfill these requirements. A simple approach is superimposing $\mathrm{LG}_p^0$ modes with certain weights so that the polynomial factors become $r^{2p_m} - {r_0}^{2p_m}$, where $p_m$ is an integer that satisfies $p_m > \left| \ell \right|/2$ and $r_0$ is the radius of the spin-down circle in the transverse plane. The corresponding beams characterized by $r^{2p_m}$ and ${r_0}^{2p_m}$ can be expressed by
\begin{equation}
    \mathrm{W}_1\left(\mathbf{r},t|\lambda\right) = \sum_{p=0}^{p_m}{(-1)^p\frac{{p_m}!}{(p_m-p)!p!}\mathrm{LG}_p^0\left(\mathbf{r},t|\lambda\right)}
\end{equation}
and 
\begin{equation}
    \mathrm{W}_2\left(\mathbf{r},t|\lambda\right) = \frac{1}{p_m!}\left(\frac{2{r_0}^2}{{w_0}^2}\right)^{p_m}\mathrm{LG}_0^0\left(\mathbf{r},t|\lambda\right),
\end{equation}
respectively. Here $w_0=\sqrt{\lambda z_r/\pi}$ is the waist radius, constant for all the modes, with $z_r$ the Rayleigh range.
Similar to the case of $P=1$, a spatiotemporal hopfionic field with $P=\ell$ can be configured by assigning modes for each wavelength and polarization as
\begin{equation}
\begin{split}
    \boldsymbol{E}\left(\mathbf{r},t\right) =&
    \left[\mathrm{W}_1\left(\mathbf{r},t|\lambda_1\right) - \mathrm{W}_2\left(\mathbf{r},t|\lambda_2\right)\right]\ket{\circlearrowleft}\\
    &+ \left[\mathrm{LG}_0^\ell\left(\mathbf{r},t|\lambda_1\right)\ket{\circlearrowright} + \mathrm{LG}_0^\ell\left(\mathbf{r},t|\lambda_2\right)\right]\ket{\circlearrowright}.
\end{split}
\end{equation}

Figure~\ref{f3}(a) depicts spatiotemporal pseudospin distributions for $s_z \leq 0$ when $l=2$, $p_m = 2$, and $r_0 = 10$\,\si{\micro\meter}, other parameters such as wavelengths and waist radius are identical to the $P=1$ configuration. In agreement with our previous predictions, the isospin curves in Fig.~\ref{f3}(a), represented by the same color tone as Fig.~\ref{f2}(f1), define a closed loop that goes twice through the torus and one time around it (for the $P=2$ case). In Fig.~\ref{f3}(b) we show the trajectories of two isospin lines of the texture in (a), clearly showing the realization of the Hopf fibration. Depicted in Fig.~\ref{f3}(c) is its corresponding hopf density $\rho_{H}(\mathbf{r},\zeta)$, yielding a perfect Hopf number $Q_H=-2$. The same plots for the $P=3$ case are presented in Figs.~\ref{f3}(d) and (e), achieved by increasing the topological charge and the radial index of the modes to $l=3$ and $p_m = 3$ respectively. In contrast, tailoring the other winding number $Q$ requires more effort than simply superimposing two colors of light, but flipping the sign of $Q_H$ can be accomplished by simply swapping the two wavelengths $\lambda_1$ and $\lambda_2$. Thus, arbitrary Hopf numbers $Q_H=\dots-3,-2,\dots,2,3\dots$ can be obtained just by modifying the indices $(p,\ell)$ of the constituent scalar modes and the wavelength $\lambda$ in which they are shaped.

We emphasize that the above discussions are strictly studied at a single transverse plane $z=0$ at the beam waist, where we can ignore the effects of beam divergence and Gouy phase, in contrast to previous works in which these features are judiciously harnessed to construct Stokes hopfions along the propagation axis \cite{shen2023topological}. Our proposed space-time hopfions exist in the spatiotemporal domain $(x,y,t,z=0)$ and at a single plane at the beam waist. However, the hopfion array can propagate along $z$ maintaining its topology ($Q_H$) when the temporal modulation of the beatings dominates over the diffraction of the modes. (see SI for the evolution in real space)

Beyond one-dimensional arrays of only temporal periodicity, 3D hopfion crystals with a periodic structure in both the transverse plane and the temporal axis can be created in the far field by assembling point-like sources with tailored polarization and phase. Multiple radiators arranged in a spatially regular manner can constitute a far field with a lattice-like structure, being also possible to create optical vortex lattices by adding appropriate phase differences between these radiators~\cite{marco2024propagation}. Following this idea, by superimposing such radiator arrays with two colors and two polarizations, a three-dimensional spatiotemporal hopfion crystal can be created, as depicted in Fig.~\ref{f4}(a). The bottom panel in Fig.~\ref{f4}(a) schematically renders the corresponding source arrangement. A collection of 16 sources except the central $\ket{\circlearrowleft}$-polarized one is configured to generate a skyrmionium lattice in the far field, as shown in Fig.~\ref{f4}(b) (more details can be found in SI). In this configuration, there are four subcells within the unit supercell, containing two skyrmioniums and two anti-skyrmioniums with opposite topological features. See the unit supercell depicted in Fig.~\ref{f4}(c), in which vortex-type and saddle-type skyrmioniums appear, which will yield opposite Hopf numbers of +1 and -1 respectively. Note that additionally to these topology changes between adjacent sub-cells, a topology-preserving helicity change of $\pi$ appears between sub-cells with the same skyrmionium number. 

Following the idea of the 1D counterparts, the 3D hopfion crystal is further synthesized by generating this skyrmionium lattice at two different wavelengths ($\lambda_1$ and $\lambda_2$) so that the lattice temporally beats. Therefore, the number of sources doubles from that shown in Fig~\ref{f4}(a). We also add another bichromatic $\ket{\circlearrowleft}$-polarized source at the center of the emitters whose temporal beating is $\pi-$shifted to the rest of the lattice, yielding a counterbeat to the beating skyrmionium lattice (see SI for the detailed source arrangement). Rigorously, the position of each source relative to the central source must be scaled in proportion to its wavelength to yield an identical period in the far field for the two wavelengths. The resulting temporal spin texture is shown in Fig.~\ref{f4}(d), and 3D Hopf density is plotted in Fig. \ref{f4}(e) for a single temporal period, featuring an alternating topological structure arising from its constituent skyrmionium and anti-skyrmionium subcells. The former yields a subcell with positive Hopf density $\rho_{\rm H}$ and a resulting Hopf number of $Q_{\rm H}=+1$, while the latter yields negative $\rho_{\rm H}$ and $Q_{\rm H}=-1$. The helicity changes in anti-skyrmionium subcells highlighted in Fig.~\ref{f4}(c) do not alter the resulting Hopf number in Fig.~\ref{f4}(e).

In summary, we present a method of generating 1D hopfion crystals with a periodic beating structure along time by tailoring dichromatic vectorial structured light and demonstrate the control of arbitrary higher-order Hopf indices. Contrary to previous works constructing spatial hopfions using the diffraction features of monochromatic Laguerre-Gaussian modes, our approach operates in the spatiotemporal domain. Here, hopfions are formed by a delicate design of the temporal beating of the constituent dichromatic components and their spatial properties, with no reliance on diffraction in our design. Furthermore, we extend this concept to present a generation method of 3D spatiotemporal hopfion crystals, with periodicity in both space and time. These structures are formed in the far field from a specially designed dichromatic dipole emitter array with tailored polarization, phase, and emitter arrangement. While we demonstrate a design of a square-lattice 3D hopfion crystal, it is theoretically feasible to achieve various types of topologies through proper manipulation of the features of the emitter array. We hope our theoretical design will stimulate the corresponding experimental observations of space-time hopfions across optical, terahertz, and microwave domains. The birth of space-time hopfion crystals opens up an exciting field of high-dimensional topological quasiparticle crystals in both condensed materials and photonic fields. This advancement promises to uncover unexplored phenomena that may arise in diverse fields such as atom trapping, light-matter interaction, topological optical communications, etc.

\begin{acknowledgments}
W. Lin and S. Iwamoto acknowledge the support from JST, CREST Grant Number JPMJCR19T1, Japan.
Y. Shen acknowledges the support from Nanyang Technological University Start Up Grant, Singapore Ministry of Education (MOE) AcRF Tier 1 grant (RG157/23), MoE AcRF Tier 1 Thematic grant (RT11/23), and Imperial-Nanyang Technological University Collaboration Fund (INCF-2024-007)
\end{acknowledgments}

\bibliography{ms}

%apsrev4-2.bst 2019-01-14 (MD) hand-edited version of apsrev4-1.bst
%Control: key (0)
%Control: author (8) initials jnrlst
%Control: editor formatted (1) identically to author
%Control: production of article title (0) allowed
%Control: page (0) single
%Control: year (1) truncated
%Control: production of eprint (0) enabled
\begin{thebibliography}{39}%
\makeatletter
\providecommand \@ifxundefined [1]{%
 \@ifx{#1\undefined}
}%
\providecommand \@ifnum [1]{%
 \ifnum #1\expandafter \@firstoftwo
 \else \expandafter \@secondoftwo
 \fi
}%
\providecommand \@ifx [1]{%
 \ifx #1\expandafter \@firstoftwo
 \else \expandafter \@secondoftwo
 \fi
}%
\providecommand \natexlab [1]{#1}%
\providecommand \enquote  [1]{``#1''}%
\providecommand \bibnamefont  [1]{#1}%
\providecommand \bibfnamefont [1]{#1}%
\providecommand \citenamefont [1]{#1}%
\providecommand \href@noop [0]{\@secondoftwo}%
\providecommand \href [0]{\begingroup \@sanitize@url \@href}%
\providecommand \@href[1]{\@@startlink{#1}\@@href}%
\providecommand \@@href[1]{\endgroup#1\@@endlink}%
\providecommand \@sanitize@url [0]{\catcode `\\12\catcode `\$12\catcode `\&12\catcode `\#12\catcode `\^12\catcode `\_12\catcode `\%12\relax}%
\providecommand \@@startlink[1]{}%
\providecommand \@@endlink[0]{}%
\providecommand \url  [0]{\begingroup\@sanitize@url \@url }%
\providecommand \@url [1]{\endgroup\@href {#1}{\urlprefix }}%
\providecommand \urlprefix  [0]{URL }%
\providecommand \Eprint [0]{\href }%
\providecommand \doibase [0]{https://doi.org/}%
\providecommand \selectlanguage [0]{\@gobble}%
\providecommand \bibinfo  [0]{\@secondoftwo}%
\providecommand \bibfield  [0]{\@secondoftwo}%
\providecommand \translation [1]{[#1]}%
\providecommand \BibitemOpen [0]{}%
\providecommand \bibitemStop [0]{}%
\providecommand \bibitemNoStop [0]{.\EOS\space}%
\providecommand \EOS [0]{\spacefactor3000\relax}%
\providecommand \BibitemShut  [1]{\csname bibitem#1\endcsname}%
\let\auto@bib@innerbib\@empty
%</preamble>
\bibitem [{\citenamefont {Bogdanov}\ and\ \citenamefont {Panagopoulos}(2020)}]{bogdanov2020physical}%
  \BibitemOpen
  \bibfield  {author} {\bibinfo {author} {\bibfnamefont {A.~N.}\ \bibnamefont {Bogdanov}}\ and\ \bibinfo {author} {\bibfnamefont {C.}~\bibnamefont {Panagopoulos}},\ }\bibfield  {title} {\bibinfo {title} {Physical foundations and basic properties of magnetic skyrmions},\ }\href {https://doi.org/10.1038/s42254-020-0203-7} {\bibfield  {journal} {\bibinfo  {journal} {Nature Reviews Physics}\ }\textbf {\bibinfo {volume} {2}},\ \bibinfo {pages} {492} (\bibinfo {year} {2020})}\BibitemShut {NoStop}%
\bibitem [{\citenamefont {G\"{o}bel}\ \emph {et~al.}(2021)\citenamefont {G\"{o}bel}, \citenamefont {Mertig},\ and\ \citenamefont {Tretiakov}}]{gobel2021beyond}%
  \BibitemOpen
  \bibfield  {author} {\bibinfo {author} {\bibfnamefont {B.}~\bibnamefont {G\"{o}bel}}, \bibinfo {author} {\bibfnamefont {I.}~\bibnamefont {Mertig}},\ and\ \bibinfo {author} {\bibfnamefont {O.~A.}\ \bibnamefont {Tretiakov}},\ }\bibfield  {title} {\bibinfo {title} {Beyond skyrmions: Review and perspectives of alternative magnetic quasiparticles},\ }\href {https://doi.org/https://doi.org/10.1016/j.physrep.2020.10.001} {\bibfield  {journal} {\bibinfo  {journal} {Physics Reports}\ }\textbf {\bibinfo {volume} {895}},\ \bibinfo {pages} {1} (\bibinfo {year} {2021})}\BibitemShut {NoStop}%
\bibitem [{\citenamefont {Bernevig}\ \emph {et~al.}(2022)\citenamefont {Bernevig}, \citenamefont {Felser},\ and\ \citenamefont {Beidenkopf}}]{bernevig2022progress}%
  \BibitemOpen
  \bibfield  {author} {\bibinfo {author} {\bibfnamefont {B.~A.}\ \bibnamefont {Bernevig}}, \bibinfo {author} {\bibfnamefont {C.}~\bibnamefont {Felser}},\ and\ \bibinfo {author} {\bibfnamefont {H.}~\bibnamefont {Beidenkopf}},\ }\bibfield  {title} {\bibinfo {title} {Progress and prospects in magnetic topological materials},\ }\href {https://doi.org/10.1038/s41586-021-04105-x} {\bibfield  {journal} {\bibinfo  {journal} {Nature}\ }\textbf {\bibinfo {volume} {603}},\ \bibinfo {pages} {41} (\bibinfo {year} {2022})}\BibitemShut {NoStop}%
\bibitem [{\citenamefont {Han}\ \emph {et~al.}(2022)\citenamefont {Han}, \citenamefont {Addiego}, \citenamefont {Prokhorenko}, \citenamefont {Wang}, \citenamefont {Fu}, \citenamefont {Nahas}, \citenamefont {Yan}, \citenamefont {Cai}, \citenamefont {Wei}, \citenamefont {Fang}, \citenamefont {Liu}, \citenamefont {Ji}, \citenamefont {Guo}, \citenamefont {Gu}, \citenamefont {Yang}, \citenamefont {Wang}, \citenamefont {Bellaiche}, \citenamefont {Chen}, \citenamefont {Wu}, \citenamefont {Nie},\ and\ \citenamefont {Pan}}]{han2022high}%
  \BibitemOpen
  \bibfield  {author} {\bibinfo {author} {\bibfnamefont {L.}~\bibnamefont {Han}}, \bibinfo {author} {\bibfnamefont {C.}~\bibnamefont {Addiego}}, \bibinfo {author} {\bibfnamefont {S.}~\bibnamefont {Prokhorenko}}, \bibinfo {author} {\bibfnamefont {M.}~\bibnamefont {Wang}}, \bibinfo {author} {\bibfnamefont {H.}~\bibnamefont {Fu}}, \bibinfo {author} {\bibfnamefont {Y.}~\bibnamefont {Nahas}}, \bibinfo {author} {\bibfnamefont {X.}~\bibnamefont {Yan}}, \bibinfo {author} {\bibfnamefont {S.}~\bibnamefont {Cai}}, \bibinfo {author} {\bibfnamefont {T.}~\bibnamefont {Wei}}, \bibinfo {author} {\bibfnamefont {Y.}~\bibnamefont {Fang}}, \bibinfo {author} {\bibfnamefont {H.}~\bibnamefont {Liu}}, \bibinfo {author} {\bibfnamefont {D.}~\bibnamefont {Ji}}, \bibinfo {author} {\bibfnamefont {W.}~\bibnamefont {Guo}}, \bibinfo {author} {\bibfnamefont {Z.}~\bibnamefont {Gu}}, \bibinfo {author} {\bibfnamefont {Y.}~\bibnamefont {Yang}}, \bibinfo {author} {\bibfnamefont {P.}~\bibnamefont {Wang}}, \bibinfo {author} {\bibfnamefont
  {L.}~\bibnamefont {Bellaiche}}, \bibinfo {author} {\bibfnamefont {Y.}~\bibnamefont {Chen}}, \bibinfo {author} {\bibfnamefont {D.}~\bibnamefont {Wu}}, \bibinfo {author} {\bibfnamefont {Y.}~\bibnamefont {Nie}},\ and\ \bibinfo {author} {\bibfnamefont {X.}~\bibnamefont {Pan}},\ }\bibfield  {title} {\bibinfo {title} {High-density switchable skyrmion-like polar nanodomains integrated on silicon},\ }\href {https://doi.org/10.1038/s41586-021-04338-w} {\bibfield  {journal} {\bibinfo  {journal} {Nature}\ }\textbf {\bibinfo {volume} {603}},\ \bibinfo {pages} {63} (\bibinfo {year} {2022})}\BibitemShut {NoStop}%
\bibitem [{\citenamefont {Chen}\ \emph {et~al.}(2024)\citenamefont {Chen}, \citenamefont {Lourembam}, \citenamefont {Ho}, \citenamefont {Toh}, \citenamefont {Huang}, \citenamefont {Chen}, \citenamefont {Tan}, \citenamefont {Yap}, \citenamefont {Lim}, \citenamefont {Tan}, \citenamefont {Suraj}, \citenamefont {Sim}, \citenamefont {Toh}, \citenamefont {Lim}, \citenamefont {Lim}, \citenamefont {Zhou}, \citenamefont {Chung}, \citenamefont {Lim},\ and\ \citenamefont {Soumyanarayanan}}]{chen2024all}%
  \BibitemOpen
  \bibfield  {author} {\bibinfo {author} {\bibfnamefont {S.}~\bibnamefont {Chen}}, \bibinfo {author} {\bibfnamefont {J.}~\bibnamefont {Lourembam}}, \bibinfo {author} {\bibfnamefont {P.}~\bibnamefont {Ho}}, \bibinfo {author} {\bibfnamefont {A.~K.~J.}\ \bibnamefont {Toh}}, \bibinfo {author} {\bibfnamefont {J.}~\bibnamefont {Huang}}, \bibinfo {author} {\bibfnamefont {X.}~\bibnamefont {Chen}}, \bibinfo {author} {\bibfnamefont {H.~K.}\ \bibnamefont {Tan}}, \bibinfo {author} {\bibfnamefont {S.~L.~K.}\ \bibnamefont {Yap}}, \bibinfo {author} {\bibfnamefont {R.~J.~J.}\ \bibnamefont {Lim}}, \bibinfo {author} {\bibfnamefont {H.~R.}\ \bibnamefont {Tan}}, \bibinfo {author} {\bibfnamefont {T.~S.}\ \bibnamefont {Suraj}}, \bibinfo {author} {\bibfnamefont {M.~I.}\ \bibnamefont {Sim}}, \bibinfo {author} {\bibfnamefont {Y.~T.}\ \bibnamefont {Toh}}, \bibinfo {author} {\bibfnamefont {I.}~\bibnamefont {Lim}}, \bibinfo {author} {\bibfnamefont {N.~C.~B.}\ \bibnamefont {Lim}}, \bibinfo {author} {\bibfnamefont {J.}~\bibnamefont
  {Zhou}}, \bibinfo {author} {\bibfnamefont {H.~J.}\ \bibnamefont {Chung}}, \bibinfo {author} {\bibfnamefont {S.~T.}\ \bibnamefont {Lim}},\ and\ \bibinfo {author} {\bibfnamefont {A.}~\bibnamefont {Soumyanarayanan}},\ }\bibfield  {title} {\bibinfo {title} {All-electrical skyrmionic magnetic tunnel junction},\ }\href {https://doi.org/10.1038/s41586-024-07131-7} {\bibfield  {journal} {\bibinfo  {journal} {Nature}\ }\textbf {\bibinfo {volume} {627}},\ \bibinfo {pages} {522} (\bibinfo {year} {2024})}\BibitemShut {NoStop}%
\bibitem [{\citenamefont {Shen}\ \emph {et~al.}(2024{\natexlab{a}})\citenamefont {Shen}, \citenamefont {Zhang}, \citenamefont {Shi}, \citenamefont {Du}, \citenamefont {Yuan},\ and\ \citenamefont {Zayats}}]{shen2024optical}%
  \BibitemOpen
  \bibfield  {author} {\bibinfo {author} {\bibfnamefont {Y.}~\bibnamefont {Shen}}, \bibinfo {author} {\bibfnamefont {Q.}~\bibnamefont {Zhang}}, \bibinfo {author} {\bibfnamefont {P.}~\bibnamefont {Shi}}, \bibinfo {author} {\bibfnamefont {L.}~\bibnamefont {Du}}, \bibinfo {author} {\bibfnamefont {X.}~\bibnamefont {Yuan}},\ and\ \bibinfo {author} {\bibfnamefont {A.~V.}\ \bibnamefont {Zayats}},\ }\bibfield  {title} {\bibinfo {title} {Optical skyrmions and other topological quasiparticles of light},\ }\href {https://doi.org/10.1038/s41566-023-01325-7} {\bibfield  {journal} {\bibinfo  {journal} {Nature Photonics}\ }\textbf {\bibinfo {volume} {18}},\ \bibinfo {pages} {15} (\bibinfo {year} {2024}{\natexlab{a}})}\BibitemShut {NoStop}%
\bibitem [{\citenamefont {He}\ \emph {et~al.}(2022)\citenamefont {He}, \citenamefont {Shen},\ and\ \citenamefont {Forbes}}]{he2022towards}%
  \BibitemOpen
  \bibfield  {author} {\bibinfo {author} {\bibfnamefont {C.}~\bibnamefont {He}}, \bibinfo {author} {\bibfnamefont {Y.}~\bibnamefont {Shen}},\ and\ \bibinfo {author} {\bibfnamefont {A.}~\bibnamefont {Forbes}},\ }\bibfield  {title} {\bibinfo {title} {Towards higher-dimensional structured light},\ }\href {https://doi.org/10.1038/s41377-022-00897-3} {\bibfield  {journal} {\bibinfo  {journal} {Light: Science {\&} Applications}\ }\textbf {\bibinfo {volume} {11}},\ \bibinfo {pages} {205} (\bibinfo {year} {2022})}\BibitemShut {NoStop}%
\bibitem [{\citenamefont {Wan}\ \emph {et~al.}(2023)\citenamefont {Wan}, \citenamefont {Wang}, \citenamefont {Liu}, \citenamefont {Fu},\ and\ \citenamefont {Shen}}]{wan2023ultra}%
  \BibitemOpen
  \bibfield  {author} {\bibinfo {author} {\bibfnamefont {Z.}~\bibnamefont {Wan}}, \bibinfo {author} {\bibfnamefont {H.}~\bibnamefont {Wang}}, \bibinfo {author} {\bibfnamefont {Q.}~\bibnamefont {Liu}}, \bibinfo {author} {\bibfnamefont {X.}~\bibnamefont {Fu}},\ and\ \bibinfo {author} {\bibfnamefont {Y.}~\bibnamefont {Shen}},\ }\bibfield  {title} {\bibinfo {title} {Ultra-degree-of-freedom structured light for ultracapacity information carriers},\ }\href {https://doi.org/10.1021/acsphotonics.2c01640} {\bibfield  {journal} {\bibinfo  {journal} {ACS Photonics}\ }\textbf {\bibinfo {volume} {10}},\ \bibinfo {pages} {2149} (\bibinfo {year} {2023})}\BibitemShut {NoStop}%
\bibitem [{\citenamefont {Lin}\ \emph {et~al.}(2021)\citenamefont {Lin}, \citenamefont {Ota}, \citenamefont {Arakawa},\ and\ \citenamefont {Iwamoto}}]{lin2021microcavity}%
  \BibitemOpen
  \bibfield  {author} {\bibinfo {author} {\bibfnamefont {W.}~\bibnamefont {Lin}}, \bibinfo {author} {\bibfnamefont {Y.}~\bibnamefont {Ota}}, \bibinfo {author} {\bibfnamefont {Y.}~\bibnamefont {Arakawa}},\ and\ \bibinfo {author} {\bibfnamefont {S.}~\bibnamefont {Iwamoto}},\ }\bibfield  {title} {\bibinfo {title} {Microcavity-based generation of full poincar\'e beams with arbitrary skyrmion numbers},\ }\href {https://doi.org/10.1103/PhysRevResearch.3.023055} {\bibfield  {journal} {\bibinfo  {journal} {Physical Review Research}\ }\textbf {\bibinfo {volume} {3}},\ \bibinfo {pages} {023055} (\bibinfo {year} {2021})}\BibitemShut {NoStop}%
\bibitem [{\citenamefont {Kerridge-Johns}\ \emph {et~al.}(2024)\citenamefont {Kerridge-Johns}, \citenamefont {Rao},\ and\ \citenamefont {Omatsu}}]{kerridge2024optical}%
  \BibitemOpen
  \bibfield  {author} {\bibinfo {author} {\bibfnamefont {W.~R.}\ \bibnamefont {Kerridge-Johns}}, \bibinfo {author} {\bibfnamefont {A.~S.}\ \bibnamefont {Rao}},\ and\ \bibinfo {author} {\bibfnamefont {T.}~\bibnamefont {Omatsu}},\ }\bibfield  {title} {\bibinfo {title} {Optical skyrmion laser using a wedged output coupler},\ }\href {https://doi.org/10.1364/optica.521901} {\bibfield  {journal} {\bibinfo  {journal} {Optica}\ }\textbf {\bibinfo {volume} {11}},\ \bibinfo {pages} {769} (\bibinfo {year} {2024})}\BibitemShut {NoStop}%
\bibitem [{\citenamefont {Fert}\ \emph {et~al.}(2017)\citenamefont {Fert}, \citenamefont {Reyren},\ and\ \citenamefont {Cros}}]{fert2017magnetic}%
  \BibitemOpen
  \bibfield  {author} {\bibinfo {author} {\bibfnamefont {A.}~\bibnamefont {Fert}}, \bibinfo {author} {\bibfnamefont {N.}~\bibnamefont {Reyren}},\ and\ \bibinfo {author} {\bibfnamefont {V.}~\bibnamefont {Cros}},\ }\bibfield  {title} {\bibinfo {title} {Magnetic skyrmions: advances in physics and potential applications},\ }\href {https://doi.org/10.1038/natrevmats.2017.31} {\bibfield  {journal} {\bibinfo  {journal} {Nature Reviews Materials}\ }\textbf {\bibinfo {volume} {2}},\ \bibinfo {pages} {17031} (\bibinfo {year} {2017})}\BibitemShut {NoStop}%
\bibitem [{\citenamefont {Yu}\ \emph {et~al.}(2018)\citenamefont {Yu}, \citenamefont {Koshibae}, \citenamefont {Tokunaga}, \citenamefont {Shibata}, \citenamefont {Taguchi}, \citenamefont {Nagaosa},\ and\ \citenamefont {Tokura}}]{yu2018transformation}%
  \BibitemOpen
  \bibfield  {author} {\bibinfo {author} {\bibfnamefont {X.~Z.}\ \bibnamefont {Yu}}, \bibinfo {author} {\bibfnamefont {W.}~\bibnamefont {Koshibae}}, \bibinfo {author} {\bibfnamefont {Y.}~\bibnamefont {Tokunaga}}, \bibinfo {author} {\bibfnamefont {K.}~\bibnamefont {Shibata}}, \bibinfo {author} {\bibfnamefont {Y.}~\bibnamefont {Taguchi}}, \bibinfo {author} {\bibfnamefont {N.}~\bibnamefont {Nagaosa}},\ and\ \bibinfo {author} {\bibfnamefont {Y.}~\bibnamefont {Tokura}},\ }\bibfield  {title} {\bibinfo {title} {Transformation between meron and skyrmion topological spin textures in a chiral magnet},\ }\href {https://doi.org/10.1038/s41586-018-0745-3} {\bibfield  {journal} {\bibinfo  {journal} {Nature}\ }\textbf {\bibinfo {volume} {564}},\ \bibinfo {pages} {95} (\bibinfo {year} {2018})}\BibitemShut {NoStop}%
\bibitem [{\citenamefont {Foster}\ \emph {et~al.}(2019)\citenamefont {Foster}, \citenamefont {Kind}, \citenamefont {Ackerman}, \citenamefont {Tai}, \citenamefont {Dennis},\ and\ \citenamefont {Smalyukh}}]{foster2019two}%
  \BibitemOpen
  \bibfield  {author} {\bibinfo {author} {\bibfnamefont {D.}~\bibnamefont {Foster}}, \bibinfo {author} {\bibfnamefont {C.}~\bibnamefont {Kind}}, \bibinfo {author} {\bibfnamefont {P.~J.}\ \bibnamefont {Ackerman}}, \bibinfo {author} {\bibfnamefont {J.-S.~B.}\ \bibnamefont {Tai}}, \bibinfo {author} {\bibfnamefont {M.~R.}\ \bibnamefont {Dennis}},\ and\ \bibinfo {author} {\bibfnamefont {I.~I.}\ \bibnamefont {Smalyukh}},\ }\bibfield  {title} {\bibinfo {title} {Two-dimensional skyrmion bags in liquid crystals and ferromagnets},\ }\href {https://doi.org/10.1038/s41567-019-0476-x} {\bibfield  {journal} {\bibinfo  {journal} {Nature Physics}\ }\textbf {\bibinfo {volume} {15}},\ \bibinfo {pages} {655} (\bibinfo {year} {2019})}\BibitemShut {NoStop}%
\bibitem [{\citenamefont {Tang}\ \emph {et~al.}(2021)\citenamefont {Tang}, \citenamefont {Wu}, \citenamefont {Wang}, \citenamefont {Kong}, \citenamefont {Lv}, \citenamefont {Wei}, \citenamefont {Zang}, \citenamefont {Tian},\ and\ \citenamefont {Du}}]{tang2021magnetic}%
  \BibitemOpen
  \bibfield  {author} {\bibinfo {author} {\bibfnamefont {J.}~\bibnamefont {Tang}}, \bibinfo {author} {\bibfnamefont {Y.}~\bibnamefont {Wu}}, \bibinfo {author} {\bibfnamefont {W.}~\bibnamefont {Wang}}, \bibinfo {author} {\bibfnamefont {L.}~\bibnamefont {Kong}}, \bibinfo {author} {\bibfnamefont {B.}~\bibnamefont {Lv}}, \bibinfo {author} {\bibfnamefont {W.}~\bibnamefont {Wei}}, \bibinfo {author} {\bibfnamefont {J.}~\bibnamefont {Zang}}, \bibinfo {author} {\bibfnamefont {M.}~\bibnamefont {Tian}},\ and\ \bibinfo {author} {\bibfnamefont {H.}~\bibnamefont {Du}},\ }\bibfield  {title} {\bibinfo {title} {Magnetic skyrmion bundles and their current-driven dynamics},\ }\href {https://doi.org/10.1038/s41565-021-00954-9} {\bibfield  {journal} {\bibinfo  {journal} {Nature Nanotechnology}\ }\textbf {\bibinfo {volume} {16}},\ \bibinfo {pages} {1086} (\bibinfo {year} {2021})}\BibitemShut {NoStop}%
\bibitem [{\citenamefont {Hayami}\ \emph {et~al.}(2021)\citenamefont {Hayami}, \citenamefont {Okubo},\ and\ \citenamefont {Motome}}]{hayami2021phase}%
  \BibitemOpen
  \bibfield  {author} {\bibinfo {author} {\bibfnamefont {S.}~\bibnamefont {Hayami}}, \bibinfo {author} {\bibfnamefont {T.}~\bibnamefont {Okubo}},\ and\ \bibinfo {author} {\bibfnamefont {Y.}~\bibnamefont {Motome}},\ }\bibfield  {title} {\bibinfo {title} {Phase shift in skyrmion crystals},\ }\href {https://doi.org/10.1038/s41467-021-27083-0} {\bibfield  {journal} {\bibinfo  {journal} {Nature Communications}\ }\textbf {\bibinfo {volume} {12}},\ \bibinfo {pages} {6927} (\bibinfo {year} {2021})}\BibitemShut {NoStop}%
\bibitem [{\citenamefont {Tsesses}\ \emph {et~al.}(2018)\citenamefont {Tsesses}, \citenamefont {Ostrovsky}, \citenamefont {Cohen}, \citenamefont {Gjonaj}, \citenamefont {Lindner},\ and\ \citenamefont {Bartal}}]{tsesses2018optical}%
  \BibitemOpen
  \bibfield  {author} {\bibinfo {author} {\bibfnamefont {S.}~\bibnamefont {Tsesses}}, \bibinfo {author} {\bibfnamefont {E.}~\bibnamefont {Ostrovsky}}, \bibinfo {author} {\bibfnamefont {K.}~\bibnamefont {Cohen}}, \bibinfo {author} {\bibfnamefont {B.}~\bibnamefont {Gjonaj}}, \bibinfo {author} {\bibfnamefont {N.~H.}\ \bibnamefont {Lindner}},\ and\ \bibinfo {author} {\bibfnamefont {G.}~\bibnamefont {Bartal}},\ }\bibfield  {title} {\bibinfo {title} {Optical skyrmion lattice in evanescent electromagnetic fields},\ }\href {https://doi.org/10.1126/science.aau0227} {\bibfield  {journal} {\bibinfo  {journal} {Science}\ }\textbf {\bibinfo {volume} {361}},\ \bibinfo {pages} {993} (\bibinfo {year} {2018})}\BibitemShut {NoStop}%
\bibitem [{\citenamefont {Du}\ \emph {et~al.}(2019)\citenamefont {Du}, \citenamefont {Yang}, \citenamefont {Zayats},\ and\ \citenamefont {Yuan}}]{du2019deep}%
  \BibitemOpen
  \bibfield  {author} {\bibinfo {author} {\bibfnamefont {L.}~\bibnamefont {Du}}, \bibinfo {author} {\bibfnamefont {A.}~\bibnamefont {Yang}}, \bibinfo {author} {\bibfnamefont {A.~V.}\ \bibnamefont {Zayats}},\ and\ \bibinfo {author} {\bibfnamefont {X.}~\bibnamefont {Yuan}},\ }\bibfield  {title} {\bibinfo {title} {Deep-subwavelength features of photonic skyrmions in a confined electromagnetic field with orbital angular momentum},\ }\href {https://doi.org/10.1038/s41567-019-0487-7} {\bibfield  {journal} {\bibinfo  {journal} {Nature Physics}\ }\textbf {\bibinfo {volume} {15}},\ \bibinfo {pages} {650} (\bibinfo {year} {2019})}\BibitemShut {NoStop}%
\bibitem [{\citenamefont {Lei}\ \emph {et~al.}(2021)\citenamefont {Lei}, \citenamefont {Yang}, \citenamefont {Shi}, \citenamefont {Xie}, \citenamefont {Du}, \citenamefont {Zayats},\ and\ \citenamefont {Yuan}}]{lei2021photonic}%
  \BibitemOpen
  \bibfield  {author} {\bibinfo {author} {\bibfnamefont {X.}~\bibnamefont {Lei}}, \bibinfo {author} {\bibfnamefont {A.}~\bibnamefont {Yang}}, \bibinfo {author} {\bibfnamefont {P.}~\bibnamefont {Shi}}, \bibinfo {author} {\bibfnamefont {Z.}~\bibnamefont {Xie}}, \bibinfo {author} {\bibfnamefont {L.}~\bibnamefont {Du}}, \bibinfo {author} {\bibfnamefont {A.~V.}\ \bibnamefont {Zayats}},\ and\ \bibinfo {author} {\bibfnamefont {X.}~\bibnamefont {Yuan}},\ }\bibfield  {title} {\bibinfo {title} {Photonic spin lattices: Symmetry constraints for skyrmion and meron topologies},\ }\href {https://doi.org/10.1103/PhysRevLett.127.237403} {\bibfield  {journal} {\bibinfo  {journal} {Physical Review Letters}\ }\textbf {\bibinfo {volume} {127}},\ \bibinfo {pages} {237403} (\bibinfo {year} {2021})}\BibitemShut {NoStop}%
\bibitem [{\citenamefont {Shen}\ \emph {et~al.}(2022)\citenamefont {Shen}, \citenamefont {Mart{\'i}nez},\ and\ \citenamefont {Rosales-Guzm{\'a}n}}]{shen2022generation}%
  \BibitemOpen
  \bibfield  {author} {\bibinfo {author} {\bibfnamefont {Y.}~\bibnamefont {Shen}}, \bibinfo {author} {\bibfnamefont {E.~C.}\ \bibnamefont {Mart{\'i}nez}},\ and\ \bibinfo {author} {\bibfnamefont {C.}~\bibnamefont {Rosales-Guzm{\'a}n}},\ }\bibfield  {title} {\bibinfo {title} {Generation of optical skyrmions with tunable topological textures},\ }\href {https://doi.org/10.1021/acsphotonics.1c01703} {\bibfield  {journal} {\bibinfo  {journal} {ACS Photonics}\ }\textbf {\bibinfo {volume} {9}},\ \bibinfo {pages} {296} (\bibinfo {year} {2022})}\BibitemShut {NoStop}%
\bibitem [{\citenamefont {Ornelas}\ \emph {et~al.}(2024)\citenamefont {Ornelas}, \citenamefont {Nape}, \citenamefont {de~Mello~Koch},\ and\ \citenamefont {Forbes}}]{ornelas2024non}%
  \BibitemOpen
  \bibfield  {author} {\bibinfo {author} {\bibfnamefont {P.}~\bibnamefont {Ornelas}}, \bibinfo {author} {\bibfnamefont {I.}~\bibnamefont {Nape}}, \bibinfo {author} {\bibfnamefont {R.}~\bibnamefont {de~Mello~Koch}},\ and\ \bibinfo {author} {\bibfnamefont {A.}~\bibnamefont {Forbes}},\ }\bibfield  {title} {\bibinfo {title} {Non-local skyrmions as topologically resilient quantum entangled states of light},\ }\href {https://doi.org/10.1038/s41566-023-01360-4} {\bibfield  {journal} {\bibinfo  {journal} {Nature Photonics}\ }\textbf {\bibinfo {volume} {18}},\ \bibinfo {pages} {258} (\bibinfo {year} {2024})}\BibitemShut {NoStop}%
\bibitem [{\citenamefont {Zheng}\ \emph {et~al.}(2023)\citenamefont {Zheng}, \citenamefont {Kiselev}, \citenamefont {Rybakov}, \citenamefont {Yang}, \citenamefont {Shi}, \citenamefont {Bl{\"u}gel},\ and\ \citenamefont {Dunin-Borkowski}}]{zheng2023hopfion}%
  \BibitemOpen
  \bibfield  {author} {\bibinfo {author} {\bibfnamefont {F.}~\bibnamefont {Zheng}}, \bibinfo {author} {\bibfnamefont {N.~S.}\ \bibnamefont {Kiselev}}, \bibinfo {author} {\bibfnamefont {F.~N.}\ \bibnamefont {Rybakov}}, \bibinfo {author} {\bibfnamefont {L.}~\bibnamefont {Yang}}, \bibinfo {author} {\bibfnamefont {W.}~\bibnamefont {Shi}}, \bibinfo {author} {\bibfnamefont {S.}~\bibnamefont {Bl{\"u}gel}},\ and\ \bibinfo {author} {\bibfnamefont {R.~E.}\ \bibnamefont {Dunin-Borkowski}},\ }\bibfield  {title} {\bibinfo {title} {Hopfion rings in a cubic chiral magnet},\ }\href {https://doi.org/10.1038/s41586-023-06658-5} {\bibfield  {journal} {\bibinfo  {journal} {Nature}\ }\textbf {\bibinfo {volume} {623}},\ \bibinfo {pages} {718} (\bibinfo {year} {2023})}\BibitemShut {NoStop}%
\bibitem [{\citenamefont {Liu}\ \emph {et~al.}(2020)\citenamefont {Liu}, \citenamefont {Hou}, \citenamefont {Han},\ and\ \citenamefont {Zang}}]{liu2020three}%
  \BibitemOpen
  \bibfield  {author} {\bibinfo {author} {\bibfnamefont {Y.}~\bibnamefont {Liu}}, \bibinfo {author} {\bibfnamefont {W.}~\bibnamefont {Hou}}, \bibinfo {author} {\bibfnamefont {X.}~\bibnamefont {Han}},\ and\ \bibinfo {author} {\bibfnamefont {J.}~\bibnamefont {Zang}},\ }\bibfield  {title} {\bibinfo {title} {Three-dimensional dynamics of a magnetic hopfion driven by spin transfer torque},\ }\href {https://doi.org/10.1103/PhysRevLett.124.127204} {\bibfield  {journal} {\bibinfo  {journal} {Physical Review Letters}\ }\textbf {\bibinfo {volume} {124}},\ \bibinfo {pages} {127204} (\bibinfo {year} {2020})}\BibitemShut {NoStop}%
\bibitem [{\citenamefont {Liu}\ \emph {et~al.}(2022)\citenamefont {Liu}, \citenamefont {Watanabe},\ and\ \citenamefont {Nagaosa}}]{liu2022emergent}%
  \BibitemOpen
  \bibfield  {author} {\bibinfo {author} {\bibfnamefont {Y.}~\bibnamefont {Liu}}, \bibinfo {author} {\bibfnamefont {H.}~\bibnamefont {Watanabe}},\ and\ \bibinfo {author} {\bibfnamefont {N.}~\bibnamefont {Nagaosa}},\ }\bibfield  {title} {\bibinfo {title} {Emergent magnetomultipoles and nonlinear responses of a magnetic hopfion},\ }\href {https://doi.org/10.1103/PhysRevLett.129.267201} {\bibfield  {journal} {\bibinfo  {journal} {Physical Review Letters}\ }\textbf {\bibinfo {volume} {129}},\ \bibinfo {pages} {267201} (\bibinfo {year} {2022})}\BibitemShut {NoStop}%
\bibitem [{\citenamefont {Kent}\ \emph {et~al.}(2021)\citenamefont {Kent}, \citenamefont {Reynolds}, \citenamefont {Raftrey}, \citenamefont {Campbell}, \citenamefont {Virasawmy}, \citenamefont {Dhuey}, \citenamefont {Chopdekar}, \citenamefont {Hierro-Rodriguez}, \citenamefont {Sorrentino}, \citenamefont {Pereiro}, \citenamefont {Ferrer}, \citenamefont {Hellman}, \citenamefont {Sutcliffe},\ and\ \citenamefont {Fischer}}]{kent2021creation}%
  \BibitemOpen
  \bibfield  {author} {\bibinfo {author} {\bibfnamefont {N.}~\bibnamefont {Kent}}, \bibinfo {author} {\bibfnamefont {N.}~\bibnamefont {Reynolds}}, \bibinfo {author} {\bibfnamefont {D.}~\bibnamefont {Raftrey}}, \bibinfo {author} {\bibfnamefont {I.~T.~G.}\ \bibnamefont {Campbell}}, \bibinfo {author} {\bibfnamefont {S.}~\bibnamefont {Virasawmy}}, \bibinfo {author} {\bibfnamefont {S.}~\bibnamefont {Dhuey}}, \bibinfo {author} {\bibfnamefont {R.~V.}\ \bibnamefont {Chopdekar}}, \bibinfo {author} {\bibfnamefont {A.}~\bibnamefont {Hierro-Rodriguez}}, \bibinfo {author} {\bibfnamefont {A.}~\bibnamefont {Sorrentino}}, \bibinfo {author} {\bibfnamefont {E.}~\bibnamefont {Pereiro}}, \bibinfo {author} {\bibfnamefont {S.}~\bibnamefont {Ferrer}}, \bibinfo {author} {\bibfnamefont {F.}~\bibnamefont {Hellman}}, \bibinfo {author} {\bibfnamefont {P.}~\bibnamefont {Sutcliffe}},\ and\ \bibinfo {author} {\bibfnamefont {P.}~\bibnamefont {Fischer}},\ }\bibfield  {title} {\bibinfo {title} {Creation and observation of hopfions in magnetic
  multilayer systems},\ }\href {https://doi.org/10.1038/s41467-021-21846-5} {\bibfield  {journal} {\bibinfo  {journal} {Nature Communications}\ }\textbf {\bibinfo {volume} {12}},\ \bibinfo {pages} {1562} (\bibinfo {year} {2021})}\BibitemShut {NoStop}%
\bibitem [{\citenamefont {Rybakov}\ \emph {et~al.}(2022)\citenamefont {Rybakov}, \citenamefont {Kiselev}, \citenamefont {Borisov}, \citenamefont {D\"{o}ring}, \citenamefont {Melcher},\ and\ \citenamefont {Bl\"{u}gel}}]{rybakov2022magnetic}%
  \BibitemOpen
  \bibfield  {author} {\bibinfo {author} {\bibfnamefont {F.~N.}\ \bibnamefont {Rybakov}}, \bibinfo {author} {\bibfnamefont {N.~S.}\ \bibnamefont {Kiselev}}, \bibinfo {author} {\bibfnamefont {A.~B.}\ \bibnamefont {Borisov}}, \bibinfo {author} {\bibfnamefont {L.}~\bibnamefont {D\"{o}ring}}, \bibinfo {author} {\bibfnamefont {C.}~\bibnamefont {Melcher}},\ and\ \bibinfo {author} {\bibfnamefont {S.}~\bibnamefont {Bl\"{u}gel}},\ }\bibfield  {title} {\bibinfo {title} {Magnetic hopfions in solids},\ }\href {https://doi.org/10.1063/5.0099942} {\bibfield  {journal} {\bibinfo  {journal} {APL Materials}\ }\textbf {\bibinfo {volume} {10}},\ \bibinfo {pages} {111113} (\bibinfo {year} {2022})}\BibitemShut {NoStop}%
\bibitem [{\citenamefont {Sugic}\ \emph {et~al.}(2021)\citenamefont {Sugic}, \citenamefont {Droop}, \citenamefont {Otte}, \citenamefont {Ehrmanntraut}, \citenamefont {Nori}, \citenamefont {Ruostekoski}, \citenamefont {Denz},\ and\ \citenamefont {Dennis}}]{sugic2021particle}%
  \BibitemOpen
  \bibfield  {author} {\bibinfo {author} {\bibfnamefont {D.}~\bibnamefont {Sugic}}, \bibinfo {author} {\bibfnamefont {R.}~\bibnamefont {Droop}}, \bibinfo {author} {\bibfnamefont {E.}~\bibnamefont {Otte}}, \bibinfo {author} {\bibfnamefont {D.}~\bibnamefont {Ehrmanntraut}}, \bibinfo {author} {\bibfnamefont {F.}~\bibnamefont {Nori}}, \bibinfo {author} {\bibfnamefont {J.}~\bibnamefont {Ruostekoski}}, \bibinfo {author} {\bibfnamefont {C.}~\bibnamefont {Denz}},\ and\ \bibinfo {author} {\bibfnamefont {M.~R.}\ \bibnamefont {Dennis}},\ }\bibfield  {title} {\bibinfo {title} {Particle-like topologies in light},\ }\href {https://doi.org/10.1038/s41467-021-26171-5} {\bibfield  {journal} {\bibinfo  {journal} {Nature Communications}\ }\textbf {\bibinfo {volume} {12}},\ \bibinfo {pages} {6785} (\bibinfo {year} {2021})}\BibitemShut {NoStop}%
\bibitem [{\citenamefont {Shen}\ \emph {et~al.}(2023{\natexlab{a}})\citenamefont {Shen}, \citenamefont {Yu}, \citenamefont {Wu}, \citenamefont {Li}, \citenamefont {Zhu},\ and\ \citenamefont {Zayats}}]{shen2023topological}%
  \BibitemOpen
  \bibfield  {author} {\bibinfo {author} {\bibfnamefont {Y.}~\bibnamefont {Shen}}, \bibinfo {author} {\bibfnamefont {B.}~\bibnamefont {Yu}}, \bibinfo {author} {\bibfnamefont {H.}~\bibnamefont {Wu}}, \bibinfo {author} {\bibfnamefont {C.}~\bibnamefont {Li}}, \bibinfo {author} {\bibfnamefont {Z.}~\bibnamefont {Zhu}},\ and\ \bibinfo {author} {\bibfnamefont {A.~V.}\ \bibnamefont {Zayats}},\ }\bibfield  {title} {\bibinfo {title} {Topological transformation and free-space transport of photonic hopfions},\ }\href {https://doi.org/10.1117/1.ap.5.1.015001} {\bibfield  {journal} {\bibinfo  {journal} {Advanced Photonics}\ }\textbf {\bibinfo {volume} {5}},\ \bibinfo {pages} {015001} (\bibinfo {year} {2023}{\natexlab{a}})}\BibitemShut {NoStop}%
\bibitem [{\citenamefont {Wang}\ and\ \citenamefont {Fan}(2023)}]{wang2023photonic}%
  \BibitemOpen
  \bibfield  {author} {\bibinfo {author} {\bibfnamefont {H.}~\bibnamefont {Wang}}\ and\ \bibinfo {author} {\bibfnamefont {S.}~\bibnamefont {Fan}},\ }\bibfield  {title} {\bibinfo {title} {Photonic spin hopfions and monopole loops},\ }\href {https://doi.org/10.1103/PhysRevLett.131.263801} {\bibfield  {journal} {\bibinfo  {journal} {Physical Review Letters}\ }\textbf {\bibinfo {volume} {131}},\ \bibinfo {pages} {263801} (\bibinfo {year} {2023})}\BibitemShut {NoStop}%
\bibitem [{\citenamefont {Shen}\ \emph {et~al.}(2023{\natexlab{b}})\citenamefont {Shen}, \citenamefont {Zhan}, \citenamefont {Wright}, \citenamefont {Christodoulides}, \citenamefont {Wise}, \citenamefont {Willner}, \citenamefont {heng Zou}, \citenamefont {Zhao}, \citenamefont {Porras}, \citenamefont {Chong}, \citenamefont {Wan}, \citenamefont {Bliokh}, \citenamefont {Liao}, \citenamefont {Hern\'{a}ndez-Garc\'{\i}a}, \citenamefont {Murnane}, \citenamefont {Yessenov}, \citenamefont {Abouraddy}, \citenamefont {Wong}, \citenamefont {Go}, \citenamefont {Kumar}, \citenamefont {Guo}, \citenamefont {Fan}, \citenamefont {Papasimakis}, \citenamefont {Zheludev}, \citenamefont {Chen}, \citenamefont {Zhu}, \citenamefont {Agrawal}, \citenamefont {Mounaix}, \citenamefont {Fontaine}, \citenamefont {Carpenter}, \citenamefont {Jolly}, \citenamefont {Dorrer}, \citenamefont {Alonso}, \citenamefont {Lopez-Quintas}, \citenamefont {L\'{o}pez-Ripa}, \citenamefont {{n}igo J~Sola}, \citenamefont {Huang}, \citenamefont {Zhang},
  \citenamefont {Ruan}, \citenamefont {Dorrah}, \citenamefont {Capasso},\ and\ \citenamefont {Forbes}}]{shen2023roadmap}%
  \BibitemOpen
  \bibfield  {author} {\bibinfo {author} {\bibfnamefont {Y.}~\bibnamefont {Shen}}, \bibinfo {author} {\bibfnamefont {Q.}~\bibnamefont {Zhan}}, \bibinfo {author} {\bibfnamefont {L.~G.}\ \bibnamefont {Wright}}, \bibinfo {author} {\bibfnamefont {D.~N.}\ \bibnamefont {Christodoulides}}, \bibinfo {author} {\bibfnamefont {F.~W.}\ \bibnamefont {Wise}}, \bibinfo {author} {\bibfnamefont {A.~E.}\ \bibnamefont {Willner}}, \bibinfo {author} {\bibfnamefont {K.}~\bibnamefont {heng Zou}}, \bibinfo {author} {\bibfnamefont {Z.}~\bibnamefont {Zhao}}, \bibinfo {author} {\bibfnamefont {M.~A.}\ \bibnamefont {Porras}}, \bibinfo {author} {\bibfnamefont {A.}~\bibnamefont {Chong}}, \bibinfo {author} {\bibfnamefont {C.}~\bibnamefont {Wan}}, \bibinfo {author} {\bibfnamefont {K.~Y.}\ \bibnamefont {Bliokh}}, \bibinfo {author} {\bibfnamefont {C.-T.}\ \bibnamefont {Liao}}, \bibinfo {author} {\bibfnamefont {C.}~\bibnamefont {Hern\'{a}ndez-Garc\'{\i}a}}, \bibinfo {author} {\bibfnamefont {M.}~\bibnamefont {Murnane}}, \bibinfo {author}
  {\bibfnamefont {M.}~\bibnamefont {Yessenov}}, \bibinfo {author} {\bibfnamefont {A.~F.}\ \bibnamefont {Abouraddy}}, \bibinfo {author} {\bibfnamefont {L.~J.}\ \bibnamefont {Wong}}, \bibinfo {author} {\bibfnamefont {M.}~\bibnamefont {Go}}, \bibinfo {author} {\bibfnamefont {S.}~\bibnamefont {Kumar}}, \bibinfo {author} {\bibfnamefont {C.}~\bibnamefont {Guo}}, \bibinfo {author} {\bibfnamefont {S.}~\bibnamefont {Fan}}, \bibinfo {author} {\bibfnamefont {N.}~\bibnamefont {Papasimakis}}, \bibinfo {author} {\bibfnamefont {N.~I.}\ \bibnamefont {Zheludev}}, \bibinfo {author} {\bibfnamefont {L.}~\bibnamefont {Chen}}, \bibinfo {author} {\bibfnamefont {W.}~\bibnamefont {Zhu}}, \bibinfo {author} {\bibfnamefont {A.}~\bibnamefont {Agrawal}}, \bibinfo {author} {\bibfnamefont {M.}~\bibnamefont {Mounaix}}, \bibinfo {author} {\bibfnamefont {N.~K.}\ \bibnamefont {Fontaine}}, \bibinfo {author} {\bibfnamefont {J.}~\bibnamefont {Carpenter}}, \bibinfo {author} {\bibfnamefont {S.~W.}\ \bibnamefont {Jolly}}, \bibinfo {author}
  {\bibfnamefont {C.}~\bibnamefont {Dorrer}}, \bibinfo {author} {\bibfnamefont {B.}~\bibnamefont {Alonso}}, \bibinfo {author} {\bibfnamefont {I.}~\bibnamefont {Lopez-Quintas}}, \bibinfo {author} {\bibfnamefont {M.}~\bibnamefont {L\'{o}pez-Ripa}}, \bibinfo {author} {\bibfnamefont {I.}~\bibnamefont {{n}igo J~Sola}}, \bibinfo {author} {\bibfnamefont {J.}~\bibnamefont {Huang}}, \bibinfo {author} {\bibfnamefont {H.}~\bibnamefont {Zhang}}, \bibinfo {author} {\bibfnamefont {Z.}~\bibnamefont {Ruan}}, \bibinfo {author} {\bibfnamefont {A.~H.}\ \bibnamefont {Dorrah}}, \bibinfo {author} {\bibfnamefont {F.}~\bibnamefont {Capasso}},\ and\ \bibinfo {author} {\bibfnamefont {A.}~\bibnamefont {Forbes}},\ }\bibfield  {title} {\bibinfo {title} {Roadmap on spatiotemporal light fields},\ }\href {https://doi.org/10.1088/2040-8986/ace4dc} {\bibfield  {journal} {\bibinfo  {journal} {Journal of Optics}\ }\textbf {\bibinfo {volume} {25}},\ \bibinfo {pages} {093001} (\bibinfo {year} {2023}{\natexlab{b}})}\BibitemShut {NoStop}%
\bibitem [{\citenamefont {Shen}\ \emph {et~al.}(2021)\citenamefont {Shen}, \citenamefont {Hou}, \citenamefont {Papasimakis},\ and\ \citenamefont {Zheludev}}]{shen2021supertoroidal}%
  \BibitemOpen
  \bibfield  {author} {\bibinfo {author} {\bibfnamefont {Y.}~\bibnamefont {Shen}}, \bibinfo {author} {\bibfnamefont {Y.}~\bibnamefont {Hou}}, \bibinfo {author} {\bibfnamefont {N.}~\bibnamefont {Papasimakis}},\ and\ \bibinfo {author} {\bibfnamefont {N.~I.}\ \bibnamefont {Zheludev}},\ }\bibfield  {title} {\bibinfo {title} {Supertoroidal light pulses as electromagnetic skyrmions propagating in free space},\ }\href {https://doi.org/10.1038/s41467-021-26037-w} {\bibfield  {journal} {\bibinfo  {journal} {Nature Communications}\ }\textbf {\bibinfo {volume} {12}},\ \bibinfo {pages} {5891} (\bibinfo {year} {2021})}\BibitemShut {NoStop}%
\bibitem [{\citenamefont {Zdagkas}\ \emph {et~al.}(2022)\citenamefont {Zdagkas}, \citenamefont {McDonnell}, \citenamefont {Deng}, \citenamefont {Shen}, \citenamefont {Li}, \citenamefont {Ellenbogen}, \citenamefont {Papasimakis},\ and\ \citenamefont {Zheludev}}]{zdagkas2022observation}%
  \BibitemOpen
  \bibfield  {author} {\bibinfo {author} {\bibfnamefont {A.}~\bibnamefont {Zdagkas}}, \bibinfo {author} {\bibfnamefont {C.}~\bibnamefont {McDonnell}}, \bibinfo {author} {\bibfnamefont {J.}~\bibnamefont {Deng}}, \bibinfo {author} {\bibfnamefont {Y.}~\bibnamefont {Shen}}, \bibinfo {author} {\bibfnamefont {G.}~\bibnamefont {Li}}, \bibinfo {author} {\bibfnamefont {T.}~\bibnamefont {Ellenbogen}}, \bibinfo {author} {\bibfnamefont {N.}~\bibnamefont {Papasimakis}},\ and\ \bibinfo {author} {\bibfnamefont {N.~I.}\ \bibnamefont {Zheludev}},\ }\bibfield  {title} {\bibinfo {title} {Observation of toroidal pulses of light},\ }\href {https://doi.org/10.1038/s41566-022-01028-5} {\bibfield  {journal} {\bibinfo  {journal} {Nature Photonics}\ }\textbf {\bibinfo {volume} {16}},\ \bibinfo {pages} {523} (\bibinfo {year} {2022})}\BibitemShut {NoStop}%
\bibitem [{\citenamefont {Wan}\ \emph {et~al.}(2022{\natexlab{a}})\citenamefont {Wan}, \citenamefont {Cao}, \citenamefont {Chen}, \citenamefont {Chong},\ and\ \citenamefont {Zhan}}]{wan2022toroidal}%
  \BibitemOpen
  \bibfield  {author} {\bibinfo {author} {\bibfnamefont {C.}~\bibnamefont {Wan}}, \bibinfo {author} {\bibfnamefont {Q.}~\bibnamefont {Cao}}, \bibinfo {author} {\bibfnamefont {J.}~\bibnamefont {Chen}}, \bibinfo {author} {\bibfnamefont {A.}~\bibnamefont {Chong}},\ and\ \bibinfo {author} {\bibfnamefont {Q.}~\bibnamefont {Zhan}},\ }\bibfield  {title} {\bibinfo {title} {Toroidal vortices of light},\ }\href {https://doi.org/10.1038/s41566-022-01013-y} {\bibfield  {journal} {\bibinfo  {journal} {Nature Photonics}\ }\textbf {\bibinfo {volume} {16}},\ \bibinfo {pages} {519} (\bibinfo {year} {2022}{\natexlab{a}})}\BibitemShut {NoStop}%
\bibitem [{\citenamefont {Wan}\ \emph {et~al.}(2022{\natexlab{b}})\citenamefont {Wan}, \citenamefont {Shen}, \citenamefont {Chong},\ and\ \citenamefont {Zhan}}]{wan2022scalar}%
  \BibitemOpen
  \bibfield  {author} {\bibinfo {author} {\bibfnamefont {C.}~\bibnamefont {Wan}}, \bibinfo {author} {\bibfnamefont {Y.}~\bibnamefont {Shen}}, \bibinfo {author} {\bibfnamefont {A.}~\bibnamefont {Chong}},\ and\ \bibinfo {author} {\bibfnamefont {Q.}~\bibnamefont {Zhan}},\ }\bibfield  {title} {\bibinfo {title} {Scalar optical hopfions},\ }\href {https://doi.org/10.1186/s43593-022-00030-2} {\bibfield  {journal} {\bibinfo  {journal} {eLight}\ }\textbf {\bibinfo {volume} {2}},\ \bibinfo {pages} {22} (\bibinfo {year} {2022}{\natexlab{b}})}\BibitemShut {NoStop}%
\bibitem [{\citenamefont {Zhan}(2024)}]{zhan2024cylindrical}%
  \BibitemOpen
  \bibfield  {author} {\bibinfo {author} {\bibfnamefont {Q.}~\bibnamefont {Zhan}},\ }\bibfield  {title} {\bibinfo {title} {Spatiotemporal sculpturing of light: a tutorial},\ }\href {https://doi.org/10.1364/aop.507558} {\bibfield  {journal} {\bibinfo  {journal} {Advances in Optics and Photonics}\ }\textbf {\bibinfo {volume} {16}},\ \bibinfo {pages} {163} (\bibinfo {year} {2024})}\BibitemShut {NoStop}%
\bibitem [{\citenamefont {Shen}\ \emph {et~al.}(2024{\natexlab{b}})\citenamefont {Shen}, \citenamefont {Papasimakis},\ and\ \citenamefont {Zheludev}}]{shen2024nondiffracting}%
  \BibitemOpen
  \bibfield  {author} {\bibinfo {author} {\bibfnamefont {Y.}~\bibnamefont {Shen}}, \bibinfo {author} {\bibfnamefont {N.}~\bibnamefont {Papasimakis}},\ and\ \bibinfo {author} {\bibfnamefont {N.~I.}\ \bibnamefont {Zheludev}},\ }\bibfield  {title} {\bibinfo {title} {Nondiffracting supertoroidal pulses and optical ``k{\'a}rm{\'a}n vortex streets''},\ }\href {https://doi.org/10.1038/s41467-024-48927-5} {\bibfield  {journal} {\bibinfo  {journal} {Nature Communications}\ }\textbf {\bibinfo {volume} {15}},\ \bibinfo {pages} {4863} (\bibinfo {year} {2024}{\natexlab{b}})}\BibitemShut {NoStop}%
\bibitem [{\citenamefont {Zhu}\ \emph {et~al.}(2024)\citenamefont {Zhu}, \citenamefont {Li}, \citenamefont {Dong}, \citenamefont {Wang}, \citenamefont {Liu}, \citenamefont {Norrman}, \citenamefont {Set\"al\"a}, \citenamefont {Cai},\ and\ \citenamefont {Chen}}]{zhu2024directional}%
  \BibitemOpen
  \bibfield  {author} {\bibinfo {author} {\bibfnamefont {X.}~\bibnamefont {Zhu}}, \bibinfo {author} {\bibfnamefont {X.}~\bibnamefont {Li}}, \bibinfo {author} {\bibfnamefont {Z.}~\bibnamefont {Dong}}, \bibinfo {author} {\bibfnamefont {F.}~\bibnamefont {Wang}}, \bibinfo {author} {\bibfnamefont {L.}~\bibnamefont {Liu}}, \bibinfo {author} {\bibfnamefont {A.}~\bibnamefont {Norrman}}, \bibinfo {author} {\bibfnamefont {T.}~\bibnamefont {Set\"al\"a}}, \bibinfo {author} {\bibfnamefont {Y.}~\bibnamefont {Cai}},\ and\ \bibinfo {author} {\bibfnamefont {Y.}~\bibnamefont {Chen}},\ }\bibfield  {title} {\bibinfo {title} {Directional difference of the spin angular-momentum density and spin vectors in tightly focused bichromatic optical lissajous beams},\ }\href {https://doi.org/10.1103/PhysRevA.109.043503} {\bibfield  {journal} {\bibinfo  {journal} {Phys. Rev. A}\ }\textbf {\bibinfo {volume} {109}},\ \bibinfo {pages} {043503} (\bibinfo {year} {2024})}\BibitemShut {NoStop}%
\bibitem [{\citenamefont {Whitehead}(1947)}]{Hopf_invariant_Whitehead}%
  \BibitemOpen
  \bibfield  {author} {\bibinfo {author} {\bibfnamefont {J.~H.~C.}\ \bibnamefont {Whitehead}},\ }\bibfield  {title} {\bibinfo {title} {An expression of hopf's invariant as an integral},\ }\href {https://doi.org/10.1073/pnas.33.5.117} {\bibfield  {journal} {\bibinfo  {journal} {Proceedings of the National Academy of Sciences}\ }\textbf {\bibinfo {volume} {33}},\ \bibinfo {pages} {117} (\bibinfo {year} {1947})}\BibitemShut {NoStop}%
\bibitem [{\citenamefont {Moore}\ \emph {et~al.}(2008)\citenamefont {Moore}, \citenamefont {Ran},\ and\ \citenamefont {Wen}}]{Hopf_calculation_Fourier}%
  \BibitemOpen
  \bibfield  {author} {\bibinfo {author} {\bibfnamefont {J.~E.}\ \bibnamefont {Moore}}, \bibinfo {author} {\bibfnamefont {Y.}~\bibnamefont {Ran}},\ and\ \bibinfo {author} {\bibfnamefont {X.-G.}\ \bibnamefont {Wen}},\ }\bibfield  {title} {\bibinfo {title} {Topological surface states in three-dimensional magnetic insulators},\ }\href {https://doi.org/10.1103/PhysRevLett.101.186805} {\bibfield  {journal} {\bibinfo  {journal} {Physical Review Letters}\ }\textbf {\bibinfo {volume} {101}},\ \bibinfo {pages} {186805} (\bibinfo {year} {2008})}\BibitemShut {NoStop}%
\bibitem [{\citenamefont {Marco}\ \emph {et~al.}(2024)\citenamefont {Marco}, \citenamefont {Herrera}, \citenamefont {Brasselet},\ and\ \citenamefont {Alonso}}]{marco2024propagation}%
  \BibitemOpen
  \bibfield  {author} {\bibinfo {author} {\bibfnamefont {D.}~\bibnamefont {Marco}}, \bibinfo {author} {\bibfnamefont {I.}~\bibnamefont {Herrera}}, \bibinfo {author} {\bibfnamefont {S.}~\bibnamefont {Brasselet}},\ and\ \bibinfo {author} {\bibfnamefont {M.~A.}\ \bibnamefont {Alonso}},\ }\href {https://doi.org/10.1021/acsphotonics.4c00292} {\bibinfo {title} {Propagation-invariant optical meron lattices}} (\bibinfo {year} {2024})\BibitemShut {NoStop}%
\end{thebibliography}%


%apsrev4-2.bst 2019-01-14 (MD) hand-edited version of apsrev4-1.bst
%Control: key (0)
%Control: author (8) initials jnrlst
%Control: editor formatted (1) identically to author
%Control: production of article title (0) allowed
%Control: page (0) single
%Control: year (1) truncated
%Control: production of eprint (0) enabled
\begin{thebibliography}{2}%
\makeatletter
\providecommand \@ifxundefined [1]{%
 \@ifx{#1\undefined}
}%
\providecommand \@ifnum [1]{%
 \ifnum #1\expandafter \@firstoftwo
 \else \expandafter \@secondoftwo
 \fi
}%
\providecommand \@ifx [1]{%
 \ifx #1\expandafter \@firstoftwo
 \else \expandafter \@secondoftwo
 \fi
}%
\providecommand \natexlab [1]{#1}%
\providecommand \enquote  [1]{``#1''}%
\providecommand \bibnamefont  [1]{#1}%
\providecommand \bibfnamefont [1]{#1}%
\providecommand \citenamefont [1]{#1}%
\providecommand \href@noop [0]{\@secondoftwo}%
\providecommand \href [0]{\begingroup \@sanitize@url \@href}%
\providecommand \@href[1]{\@@startlink{#1}\@@href}%
\providecommand \@@href[1]{\endgroup#1\@@endlink}%
\providecommand \@sanitize@url [0]{\catcode `\\12\catcode `\$12\catcode `\&12\catcode `\#12\catcode `\^12\catcode `\_12\catcode `\%12\relax}%
\providecommand \@@startlink[1]{}%
\providecommand \@@endlink[0]{}%
\providecommand \url  [0]{\begingroup\@sanitize@url \@url }%
\providecommand \@url [1]{\endgroup\@href {#1}{\urlprefix }}%
\providecommand \urlprefix  [0]{URL }%
\providecommand \Eprint [0]{\href }%
\providecommand \doibase [0]{https://doi.org/}%
\providecommand \selectlanguage [0]{\@gobble}%
\providecommand \bibinfo  [0]{\@secondoftwo}%
\providecommand \bibfield  [0]{\@secondoftwo}%
\providecommand \translation [1]{[#1]}%
\providecommand \BibitemOpen [0]{}%
\providecommand \bibitemStop [0]{}%
\providecommand \bibitemNoStop [0]{.\EOS\space}%
\providecommand \EOS [0]{\spacefactor3000\relax}%
\providecommand \BibitemShut  [1]{\csname bibitem#1\endcsname}%
\let\auto@bib@innerbib\@empty
%</preamble>
\bibitem [{\citenamefont {Whitehead}(1947)}]{Hopf_invariant_Whitehead}%
  \BibitemOpen
  \bibfield  {author} {\bibinfo {author} {\bibfnamefont {J.~H.~C.}\ \bibnamefont {Whitehead}},\ }\bibfield  {title} {\bibinfo {title} {An expression of hopf's invariant as an integral},\ }\href {https://doi.org/10.1073/pnas.33.5.117} {\bibfield  {journal} {\bibinfo  {journal} {Proceedings of the National Academy of Sciences}\ }\textbf {\bibinfo {volume} {33}},\ \bibinfo {pages} {117} (\bibinfo {year} {1947})}\BibitemShut {NoStop}%
\bibitem [{\citenamefont {Moore}\ \emph {et~al.}(2008)\citenamefont {Moore}, \citenamefont {Ran},\ and\ \citenamefont {Wen}}]{Hopf_calculation_Fourier}%
  \BibitemOpen
  \bibfield  {author} {\bibinfo {author} {\bibfnamefont {J.~E.}\ \bibnamefont {Moore}}, \bibinfo {author} {\bibfnamefont {Y.}~\bibnamefont {Ran}},\ and\ \bibinfo {author} {\bibfnamefont {X.-G.}\ \bibnamefont {Wen}},\ }\bibfield  {title} {\bibinfo {title} {Topological surface states in three-dimensional magnetic insulators},\ }\href {https://doi.org/10.1103/PhysRevLett.101.186805} {\bibfield  {journal} {\bibinfo  {journal} {Physical Review Letters}\ }\textbf {\bibinfo {volume} {101}},\ \bibinfo {pages} {186805} (\bibinfo {year} {2008})}\BibitemShut {NoStop}%
\end{thebibliography}%

\end{document}

\title{Supplementary Information for Space-Time Hopfion Crystals}
\maketitle

\renewcommand{\thesection}{S\arabic{section}}
\renewcommand{\theequation}{S\arabic{equation}}
\renewcommand{\thefigure}{S\arabic{figure}}
\setcounter{secnumdepth}{1}
\renewcommand{\bibnumfmt}[1]{[S#1]}
\renewcommand{\citenumfont}[1]{S#1}

\section{Calculation of space-time Hopf number}

We characterize the topology of the space-time pseudospin field by calculating its topological invariant of Hopf index: 
\begin{equation}\label{eq:hopf_number}
    Q_{H}=  \frac{1}{(4\pi)^2}\iiint_{\Omega^3}{\mathbf{F}\cdot\mathbf{A}}\text{d}x\text{d}y\text{d}\zeta,
\end{equation} 
where the ``emergent magnetic field" $\mathbf{F}(\mathbf{r}_{\perp},\zeta=ct)$ is defined as $F_i=\varepsilon_{ijk}\boldsymbol{s}\cdot(\partial_j\boldsymbol{s}\times\partial_k\boldsymbol{s})/2$, in which $i,j,k\equiv\{x,y,\zeta\}$, $\varepsilon_{ijk}$ is the Levi-Civita tensor, and the integral extends through the region $\Omega^3$ of the 3D space-time where the hopfion is confined. $\mathbf{A}(\mathbf{r},\zeta)$ acts as the ``vector potential" of $\mathbf{F}(\mathbf{r},\zeta)$ satisfying the vector differential equation
\begin{equation}\label{eq:vector_potential}
    \nabla\times\mathbf{A}(\mathbf{r}_{\perp},\zeta)=\mathbf{F}(\mathbf{r}_{\perp},\zeta).
\end{equation}
The scalar field $\rho_{H}(\mathbf{r},\zeta)=\mathbf{F}(\mathbf{r},\zeta)\cdot\mathbf{A}(\mathbf{r},\zeta)$ as the integral kernel of Eq.~(\ref{eq:hopf_number}) is the resultant Hopf density of the spatiotemporal pseuodspin field $\boldsymbol{s}(\mathbf{r},\zeta)$. Note that the ``vector potential" in Eq.~(\ref{eq:vector_potential}) has infinite solutions depending on the choice of gauge, thus yielding different shapes for $\rho_{H}(\mathbf{r},\zeta)$, although its integration in $\Omega^3$ is gauge invariant and gives the same $Q_H$ as shown by Whitehead in \cite{Hopf_invariant_Whitehead}. Under the choice of Coulomb's gauge $\nabla\cdot\mathbf{A}=0$, the expression of the vector potential in momentum space can be found analytically~\cite{Hopf_calculation_Fourier}.

We solve the spatiotemporal (ST) differential equation Eq.~(\ref{eq:vector_potential}) performing ST 3D Fourier transform from the real space $(x,y,\zeta=ct)$ to ST Fourier space $\boldsymbol{\kappa}=(k_x,k_y,\nu)$, $k_x$ and $k_y$ representing transverse spatial frequencies and $\nu$ the temporal frequency. Under the Coulomb gauge $\nabla\cdot\mathbf{A}(\mathbf{r}_{\perp},\zeta)=0$, the components of the ``vector potential" $\hat{\mathbf{A}}(\boldsymbol{\kappa})$ in ST Fourier space can be expressed as
$$\hat{A}_x=-i(\nu\hat{F}_y-k_y\hat{F}_\zeta)/\kappa^2,$$
$$\hat{A}_y=-i(k_x\hat{F}_\zeta-\nu\hat{F}_x)/\kappa^2,$$
$$\hat{A}_\zeta=-i(k_y\hat{F}_x-k_x\hat{F}_y)/\kappa^2,$$
or in a more compact form in real space
\begin{equation}\label{eq:Fourier_potential}
    \mathbf{A}(\mathbf{r},\zeta)=\mathcal{F}^{-1}\left\{\frac{1}{i\kappa^2}\hat{\mathbf{F}}(\boldsymbol{\kappa})\times\boldsymbol{\kappa}\right\},
\end{equation}
with $\kappa^2=k_x^2+k_y^2+\nu^2$, $\hat{\mathbf{F}}(\boldsymbol{\kappa})=(\hat{F}_x,\hat{F}_y,\hat{F}_{\zeta})$ is the ST Fourier transform of the emergent magnetic field $\mathbf{F}(\mathbf{r}_{\perp},\zeta)$ and $\mathcal{F}^{-1}$ denotes inverse ST Fourier transform. In our calculations Eq. (\ref{eq:Fourier_potential}) is solved numerically by standard FFT algorithm with the ``magnetic field" $F_i=\varepsilon_{ijk}\boldsymbol{}{s}\cdot(\partial_j\boldsymbol{}{s}\times\partial_k\boldsymbol{}{s})/2$ defined from the ST pseudo-spin field of Eqs. (1) and (2) from the main text, and we find the Hopf density $\rho_{H}(\mathbf{r}_{\perp},\zeta)$ and its corresponding Hopf charge $Q_H$.

\section{Diffraction effects in 1D space-time hopfion crystal}
\begin{figure}[t]
    \centering
    \includegraphics[]{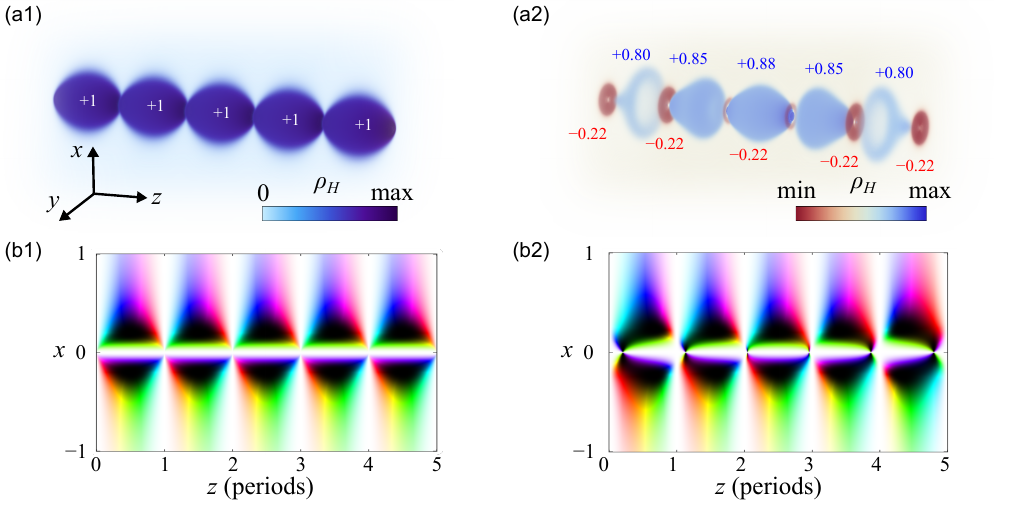}
    \caption{Flying hopfion in real XYZ space: (a) Hopfion density in XYZ and (b) its corresponding pseudo-spin texture for $y=0$ slice. In the first case (a1) and (b1) the diffraction of the modes is negligible, yielding a Hopf number of $Q_H\approx4.98$, while in the second (a2) and (b2) the constituent modes diffract and yield $Q_H\approx3.5$.}
    \label{fig:hopfion_XYZ}
\end{figure}
In the main text, we restrict the analysis to the beam waist $z=0$ and we work with the spatiotemporal coordinates $x,y,t$. This implies that the diffraction of the constituent modes does not play any role in the construction of the hopfion arrays. However, with an appropriate design of the beam parameters, the resulting hopfions can fly a certain distance and maintain their topological features within it. This is the case when the modulations introduced by the temporal envelope (beating) dominate over the diffraction of the modes, or in other words, when the Rayleigh range is large compared to the temporal (spatial) period. An example of such behavior is depicted in Fig. \ref{fig:hopfion_XYZ}-(a1), where the Rayleigh range is longer than 300 spatial periods and 5 spatial periods of the Hopf density are plotted, all of them yielding ideal Hopf number ($Q_H=+1$) after integration. However, when the spatial periodicity along $z$ is comparable with the Rayleigh range of the modes, e.g., about 3 spatial periods as depicted in (a2), the hopfion array significantly collapses, acquiring undesired negative Hopf density and yielding non-ideal Hopf numbers ($Q_H<+1$).

\section{Source arrangement for generating 3D hopfion crystal}
\begin{table*}[b]
    \centering
    \caption{Source configuration for generating a square-lattice hopfion crystal beam}
    \begin{tabular}{|c|c|c|c|c|c|}
        \hline
        Source index & $x$ position & $y$ position & Polarization & Complex amplitude of $\lambda_1$ & Complex amplitude of $\lambda_2$ \\
        \hline
        0 & $0$ & $0$ & $\circlearrowleft$ & $1$ & $-1$ \\
        1 & $2d$ & $-d$ & $\circlearrowright$ & $1$ & $1$ \\
        2 & $2d$ & $d$ & $\circlearrowright$ & $1$ & $1$ \\
        3 & $d$ & $2d$ & $\circlearrowright$ & $-i$ & $-i$ \\
        4 & $-d$ & $2d$ & $\circlearrowright$ & $-i$ & $-i$ \\
        5 & $-2d$ & $d$ & $\circlearrowright$ & $-1$ & $-1$ \\
        6 & $-2d$ & $-d$ & $\circlearrowright$ & $-1$ & $-1$ \\
        7 & $-d$ & $-2d$ & $\circlearrowright$ & $i$ & $i$ \\
        8 & $d$ & $-2d$ & $\circlearrowright$ & $i$ & $i$ \\
        9 & $2d$ & $0$ & $\circlearrowleft$ & $1$ & $1$ \\
        10 & $0$ & $2d$ & $\circlearrowleft$ & $1$ & $1$ \\
        11 & $-2d$ & $0$ & $\circlearrowleft$ & $1$ & $1$ \\
        12 & $0$ & $-2d$ & $\circlearrowleft$ & $1$ & $1$ \\
        13 & $4d$ & $0$ & $\circlearrowleft$ & $-1/2$ & $-1/2$ \\
        14 & $0$ & $4d$ & $\circlearrowleft$ & $-1/2$ & $-1/2$ \\
        15 & $-4d$ & $0$ & $\circlearrowleft$ & $-1/2$ & $-1/2$ \\
        16 & $0$ & $-4d$ & $\circlearrowleft$ & $-1/2$ & $-1/2$ \\
        \hline
    \end{tabular}
    \label{tab:S1}
\end{table*}
\begin{figure*}[tb]
    \centering
    \includegraphics[]{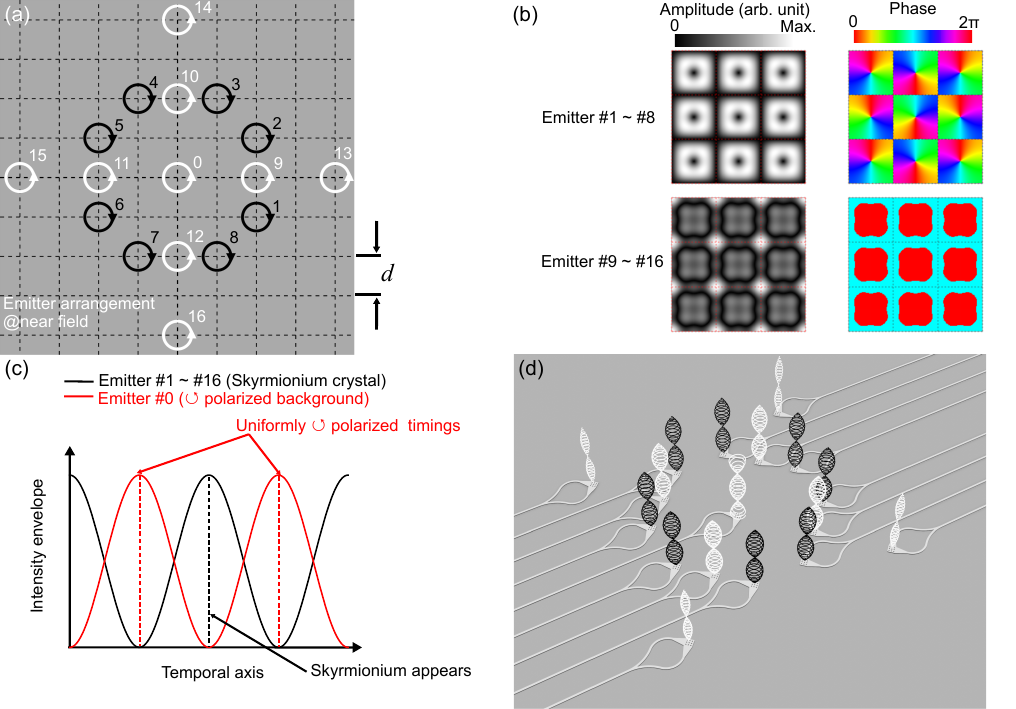}
    \caption{
        (a)~Source arrangement for generating a square-lattice hopfion crystal beam.
        (b)~Amplitude and phase distributions of far-field radiation from select radiation sources. $\ket{\circlearrowright}$ sources (emitters \#{1} to \#{8}) generate a far-field with finite OAM ($\pm1$) in each cell, while $\ket{\circlearrowleft}$ sources except the central one (emitters \#{9} to \#{16}) produce a far-field without OAM but exhibit closed-loop intensity zero in each cell. Their superposition synthesizes a skyrmionium crystal.
        (c)~Mechanism of hopfion crystal formation. When light sources \#{1} to \#{16} emit bichromatic light, their superposition creates a skyrmionium crystal beam with intensity beats along the temporal axis. Introducing an additional bichromatic light source (\#{0}) with opposite beat timing and polarization matching the skyrmionium center ($\ket{\circlearrowleft}$), skyrmionium forms only at specific temporal intervals, gradually transitioning to uniform polarization, thus yielding a hopfion crystal.
        (d)~Nanophotonic device concept implementing (a). Seventeen bifurcated grating couplers generate circularly polarized light by diffracting and superimposing orthogonal linearly polarized light with $\pm\frac{\pi}{2}$-phase shift. Illuminating each grating coupler with bichromatic light realizes the light source arrangement depicted in (a).
    }
    \label{fig:S2}
\end{figure*}
We employed a bichromatic light field generated by 17 dipole sources per wavelength to construct a three-dimensional lattice of hopfions. The emitter arrangement is illustrated in Fig.~\ref{fig:S2}(a), while the positional, polarization, and complex amplitude details of each dipole are provided in Table~\ref{tab:S1}. Our methodology entails initially generating a skyrmionium crystal and subsequently temporally modulating it using two distinct wavelengths. As depicted in Fig.~\ref{fig:S2}(b), 16 of the 17 sources (excluding the central one) are configured to produce a skyrmionium crystal, wherein the $\ket{\circlearrowright}$ and $\ket{\circlearrowleft}$ components establish an optical lattice with and without orbital angular momentum (OAM), respectively. The OAM lattice features a $\pi$-phase shift between neighboring unit cells, thereby effectuating a supercell comprising four unit cells. The momentum space period of the unit cell is $\Delta k = \frac{\pi\lambda}{d}$, necessitating the scaling of source spacing $d$ relative to the wavelength $\lambda$ to achieve identical far-field lattice constant for both colors of light. When two colors of skyrmionium crystals with aligned lattice constants are superimposed, their intensity envelopes nullify at a specific temporal interval, as illustrated in Fig.~\ref{fig:S2}(c). By overlaying an additional bichromatic light field with a finite intensity envelope at that juncture to fulfill the boundary conditions of hopfions, i.e., matching the polarization state of the skyrmionium's center ($\ket{\circlearrowleft}$), a three-dimensional crystal of hopfions is generated. A straightforward approach to meet this criterion involves adding an extra bichromatic dipole source, with one of the two wavelengths being $\pi$-phase shifted relative to the other 16 sources, as outlined in Table~\ref{tab:S1}. Such a configuration of light sources is anticipated to be feasible in the microwave domain via dipole antenna arrangements, and likewise in the optical frequency range through the utilization of spatial light modulators or dielectric micro-antenna arrays (grating couplers, as depicted in Fig.~\ref{fig:S2}(d)). Ideally, 33 radiation sources with distinct spatial positions are necessitated; however, when the wavelengths of the bichromatic light are closely aligned (resulting in a long beat period), the variation in far-field lattice constants due to wavelength discrepancies can be nearly negligible. For instance, in the context of generating beats at gigahertz frequencies within the telecommunication C-band, the wavelength disparity is sub-nanometer. Consequently, it is anticipated that numerous periods of hopfion crystals can be observed even if two wavelengths are emitted by a solitary antenna, as exemplified in Fig.~\ref{fig:S2}(d).

\section{Supplementary figures on the spatiotemporal hopfion of $Q_\mathrm{H} = -1$}
\begin{figure}[h]
    \centering
    \includegraphics{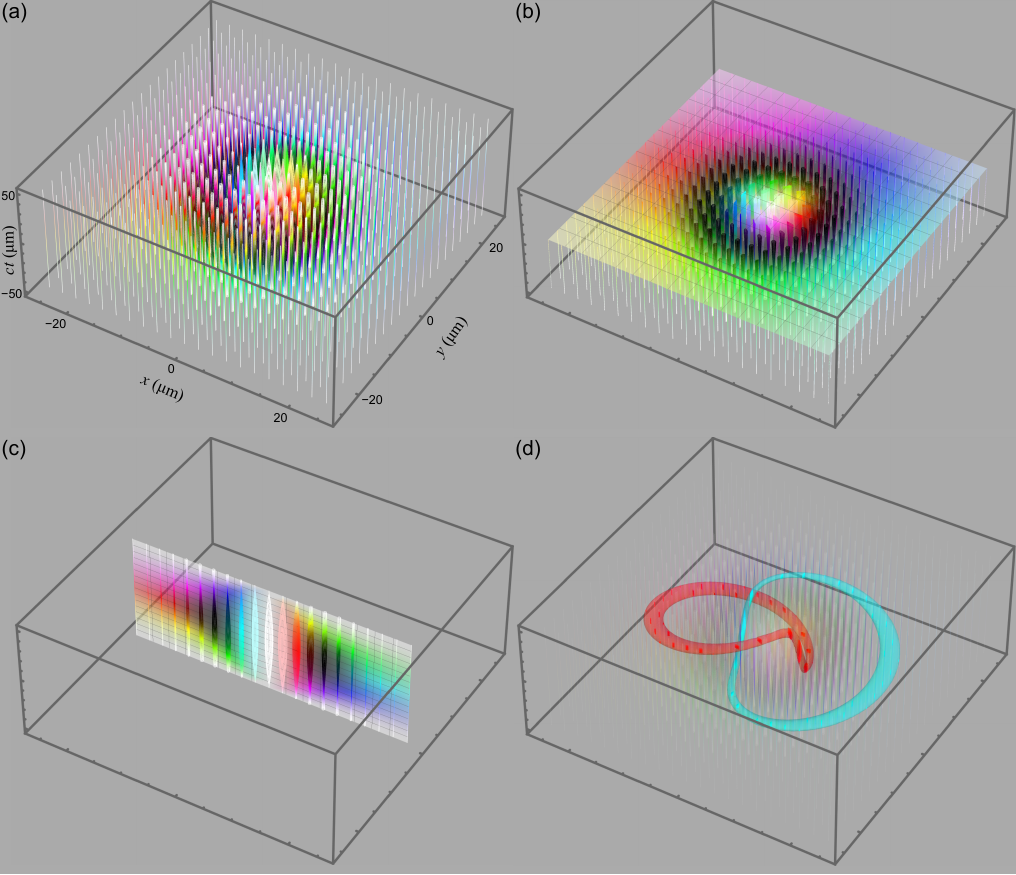}
    \caption{
        (a)~Spatiotemporal electric field trajectories of Fig.~2(f) in the main text. A single temporal period is rendered. (b)~$t=0$ slice and (c)~$y=0$ slice of (a). (d)~Distributions of horizontal polarization (the region of $s_x \geq 0.99$ is emphasized) and vertical polarization ($s_x \leq -0.99$).
    }
    \label{fig:S3}
\end{figure}

\bibliography{sm}